\documentclass{article}

\newcommand{\ken}{Sister Kenneth}
\newcommand{\ctwo}{Carmelle Zserdin, BVM}
\newcommand{\cthree}{Catherine Dunn, BVM}
\newcommand{\bone}{Bertha Fox, BVM}

\usepackage{hyperref}
\usepackage{graphicx}

\usepackage[pdftex]{color}

\begin{document}

\title{Mary Kenneth Keller: First US PhD in Computer Science}

\author{Jennifer Head
\thanks{Archivist, Sisters of Charity of the Blessed Virgin Mary, Dubuque, IA}
\and
Dianne P.~O'Leary
\thanks{Computer Science Department and Institute for Advanced Computer Studies,
University of Maryland} }

\maketitle

\begin{abstract}
In June 1965, Sister Mary Kenneth Keller, BVM, received the first US PhD in Computer Science, and this paper outlines her life and accomplishments. As a scholar, she has the distinction of being an early advocate of learning-by-example in artificial intelligence. Her main scholarly contribution was in shaping computer science education in high schools and small colleges. She was an evangelist for viewing the computer as a symbol manipulator, for providing computer literacy to everyone, and for the use of computers in service to humanity. She was far ahead of her time in working to ensure a place for women in technology and in eliminating barriers preventing their participation, such as poor access to education and daycare. She was a strong and spirited woman, a visionary in seeing how computers would revolutionize our lives. 

A condensation of this paper appeared as, ``The Legacy of Mary Kenneth Keller, First U.S. Ph.D. in Computer Science," Jennifer Head and Dianne P. O'Leary, {\em IEEE Annals of the History of Computing} 45(1):55--63, January-March 2023. \url{https://doi.org/10.1109/MAHC.2022.3231763}
\end{abstract}

\section{Introduction}

The founding of the first computer science programs at universities in the United States in the early 1960s resulted, in 1965, in the first two doctoral-level degrees. On June 7, 1965, Irving C. Tang was awarded a D.Sc. from the Applied Mathematics and Computer Science Department at Washington University in St. Louis. 
Tang's thesis was entitled ``Radial Flow Between Parallel Planes,"
directed by C. David Gorman of the Mathematics Department.
The same day \cite{LondonBlog}, Mary Kenneth Keller received a PhD in Computer Science from the University of Wisconsin in Madison. We can credit Ralph London for documenting this and correcting the historical record \cite{LondonBlog,London2}.

\begin{figure}
\begin{centering}
\includegraphics[width=.4\columnwidth] {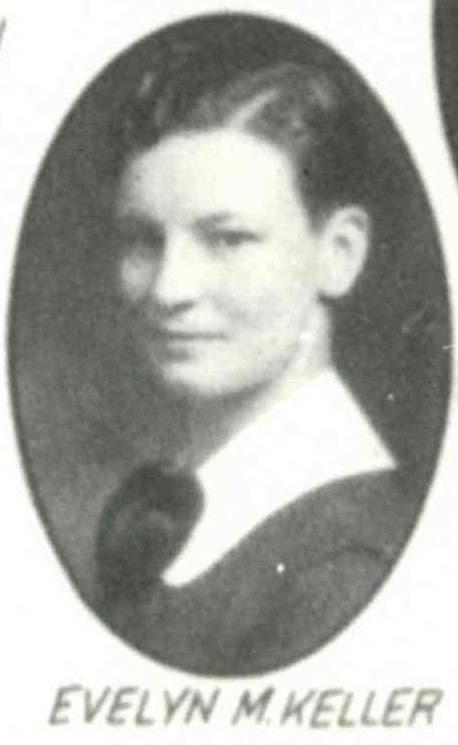}
\hspace{.05\columnwidth}
\includegraphics[width=.44\columnwidth] {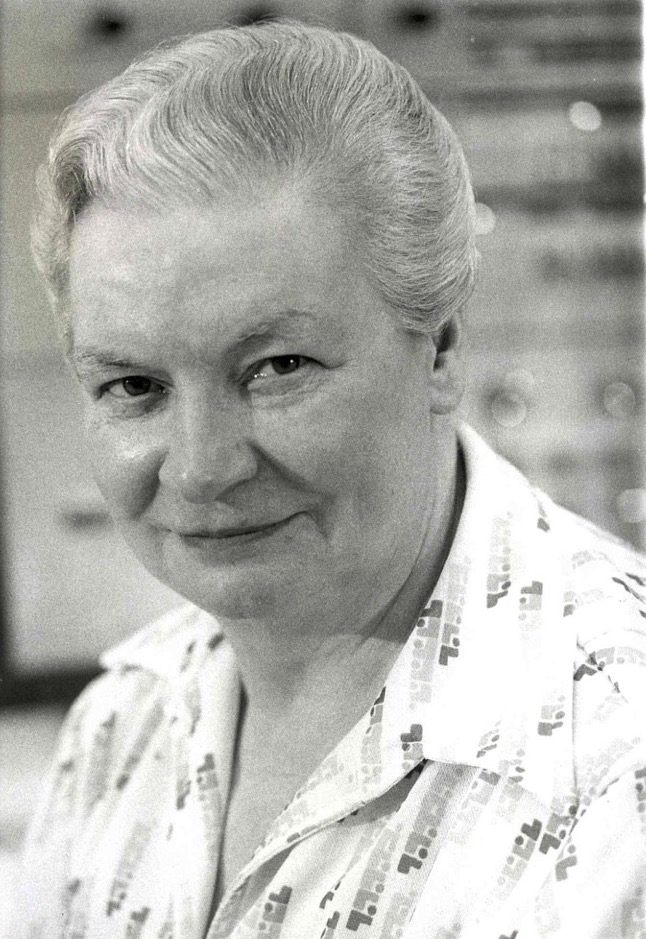}
\\
\end{centering}
\caption{\ken's high school yearbook photo and a photo from later in life. Photo credits: BVM Archives and Clarke University Archives.}
\label{fig_twophotos}
\end{figure}

In this paper, we focus on Mary Kenneth Keller ({\bf Figure \ref{fig_twophotos}}).
She was a woman of many names. Born Evelyn Marie Keller in Cleveland, Ohio, she entered the Catholic congregation of the Sisters of Charity of the Blessed Virgin Mary (BVMs) and was given the name Sister Mary Kenneth. Like the biblical Abram becoming Abraham, and Saul becoming Paul, religious sisters were given new names to signify their new mission. To her fellow sisters, she would be Kenneth or Ken, because all of them had some form of Mary in their names. To her students, she would be Sister Kenneth (or perhaps simply ``S'ter"). In her publications, she used the name Sister Mary K. Keller \cite{umap1,umap2,umap3,umap4,siguuc1,Dorn}, Sister Mary Kenneth \cite{bvm1},  Mary K. Keller \cite{curric}, and (her complete name) Sister Mary Kenneth Keller, BVM \cite{phd1}. In this paper, we refer to her as \ken.

\section{LIFE BEFORE COMPUTING}

\ken\ was born Dec 17, 1913, in Cleveland, Ohio, and the
1920 US census lists her home as  E 79th Street, Cleveland Ward 24, Cuyahoga, Ohio.
\ken\ had an older brother, Ralph L. Keller (1908-1998), 
a younger sister Ruth Catherine Keller Peters (1917-1999), 
and a younger brother,  John (Jack) Franklin Keller (1921-1987), all born in Cleveland.
By 1930 (census records), and probably around 1923 (according to a letter from her parish priest), her family had moved to  6056 N. Hermitage Avenue, Chicago. Her father, John Keller, was born in New York around 1882 and worked as an embosser in lithographing.
Her mother, Catherine Sullivan Keller, born in 1884 in Ohio, was the daughter of Irish immigrants. Both mother and father had 8th grade educations (1940 Census).
\ken\ attended St. Philip Neri Catholic Elementary School in Cleveland, taught by the Sisters of the Humility of Mary (CHMs), and St. Gertrude Catholic Elementary School in Chicago, taught by BVM sisters. 

\begin{figure}
\begin{centering}
\includegraphics[width=.97\columnwidth] {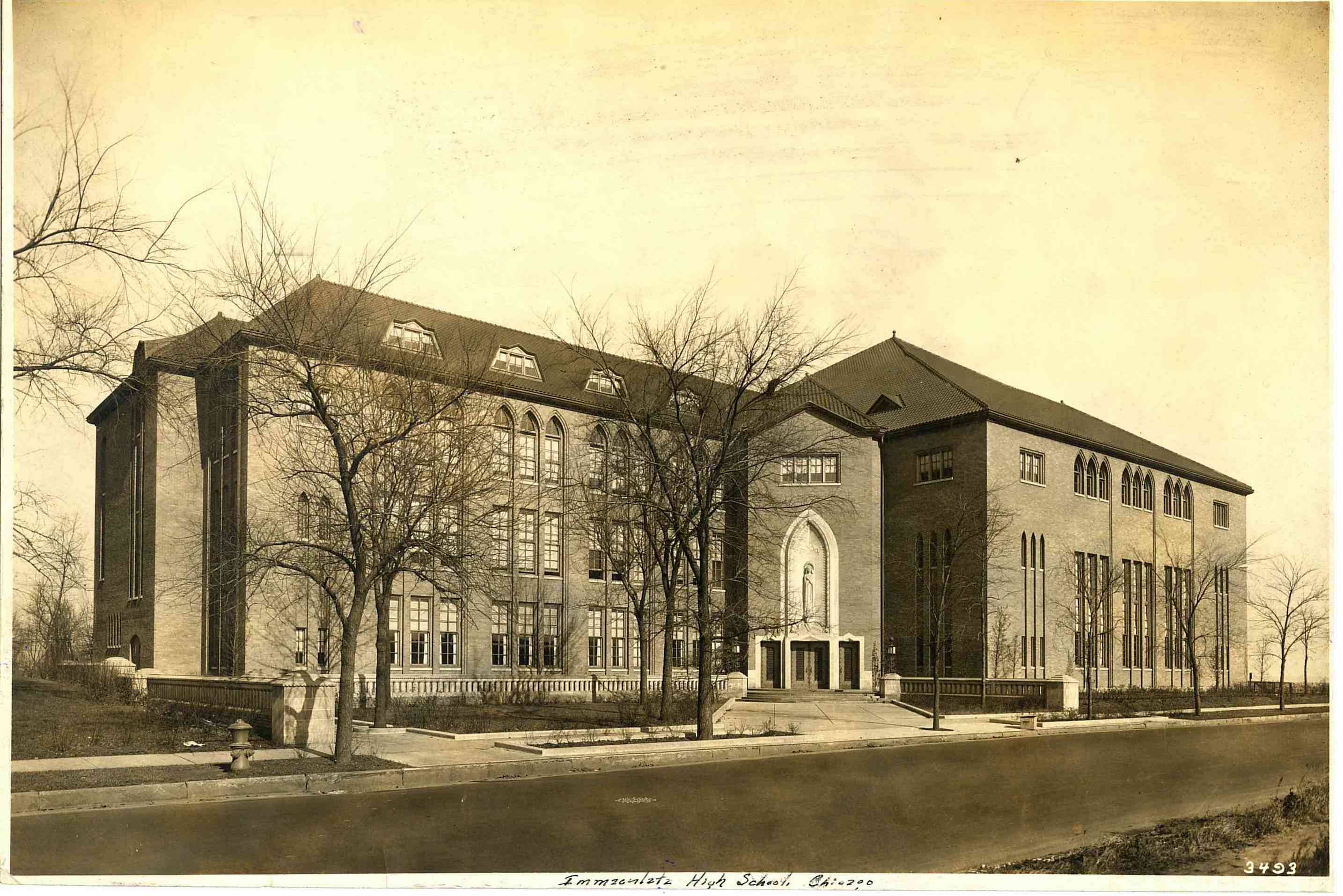}
\\
\end{centering}
\caption{The Immaculata High School, Chicago, Illinois, as it appeared in 1930. Photo credit: The BVM Archives.}
\label{fig_mac}
\end{figure}

After that, \ken\ attended The Immaculata High School, Chicago, another BVM school ({\bf Figure \ref{fig_mac}}). She was quite involved in school activities: ``Social Chairman, '30; Sec., '28; Promoter, '28, '29, '30, '31; Braille, '30, '31 (Rep.); Band, '30; Orchestra, '31 (Librarian); Basketball, '28, '29, '30, '31; News Staff, '28, '30, '31." \cite{mac193106}. She participated in intramural basketball tournaments, playing forward. She played French horn in the newly-formed school band and violin in the newly-formed orchestra. She covered music and other areas for the school newspaper; samples of her writing are given in an appendix. She was quite active in the leadership of the Braille Club, a division of the school's Sodality, promoting religious devotion and social service. Members of the Braille Club translated books into Braille to aid blind persons. In 1930, she earned one of two Braille certificates issued by the Washington, DC Braille offices \cite{mac193011}. Her life-long interest in the theater might have begun with a rather minor part as Godfrey de Bouillon, Duke of Lorraine, in the senior play, ``When All the World Was Young" \cite{mac193102}.
In 1931, she graduated. Her address at graduation was 1100 S. Hoyne Avenue \cite{Beckman}.

\begin{figure}
\begin{centering}
\includegraphics[width=.82\columnwidth] {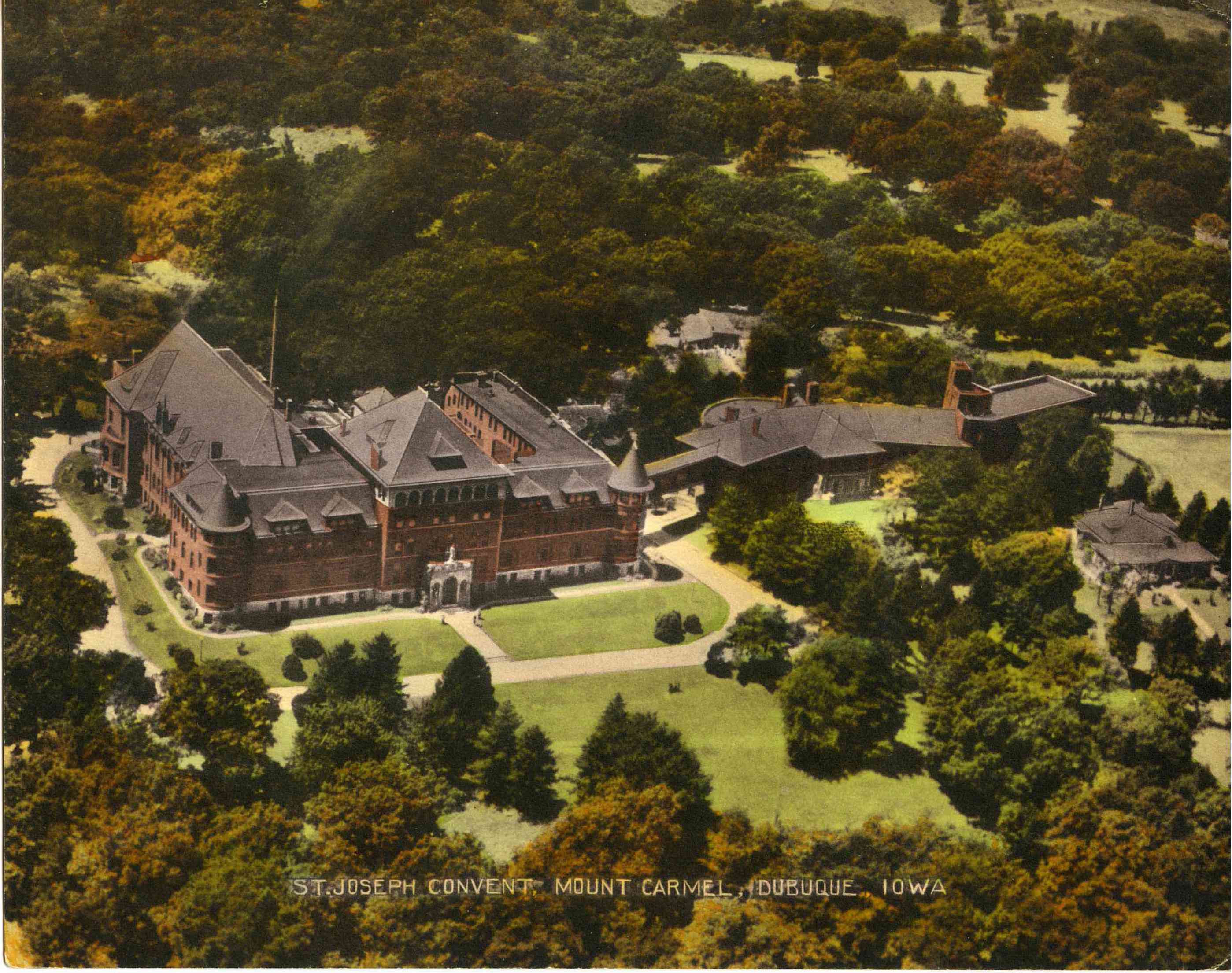}
\\
\end{centering}
\caption{The BVM motherhouse, Mount Carmel, in Dubuque, Iowa, which \ken\ entered in 1932. Photo credit: The BVM Archives.}
\label{fig_mtcarmel}
\end{figure}

\begin{figure*}
\begin{centering}
\includegraphics[width=.45\columnwidth] {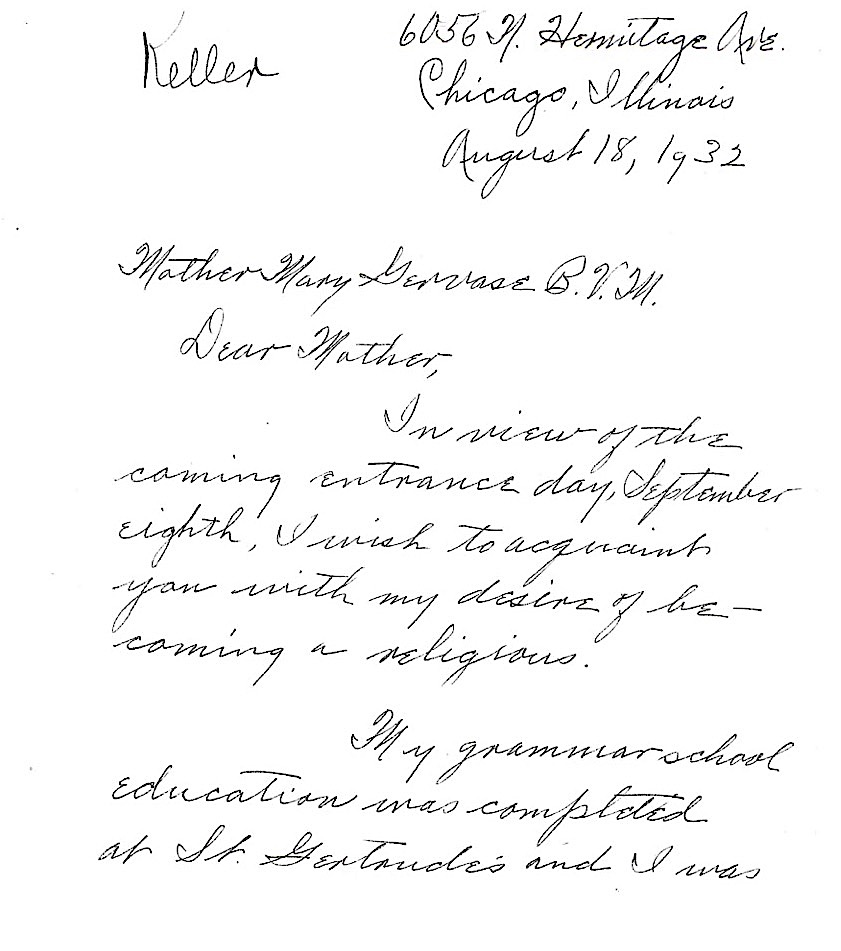}
\hspace{.05\columnwidth}
\includegraphics[width=.45\columnwidth] {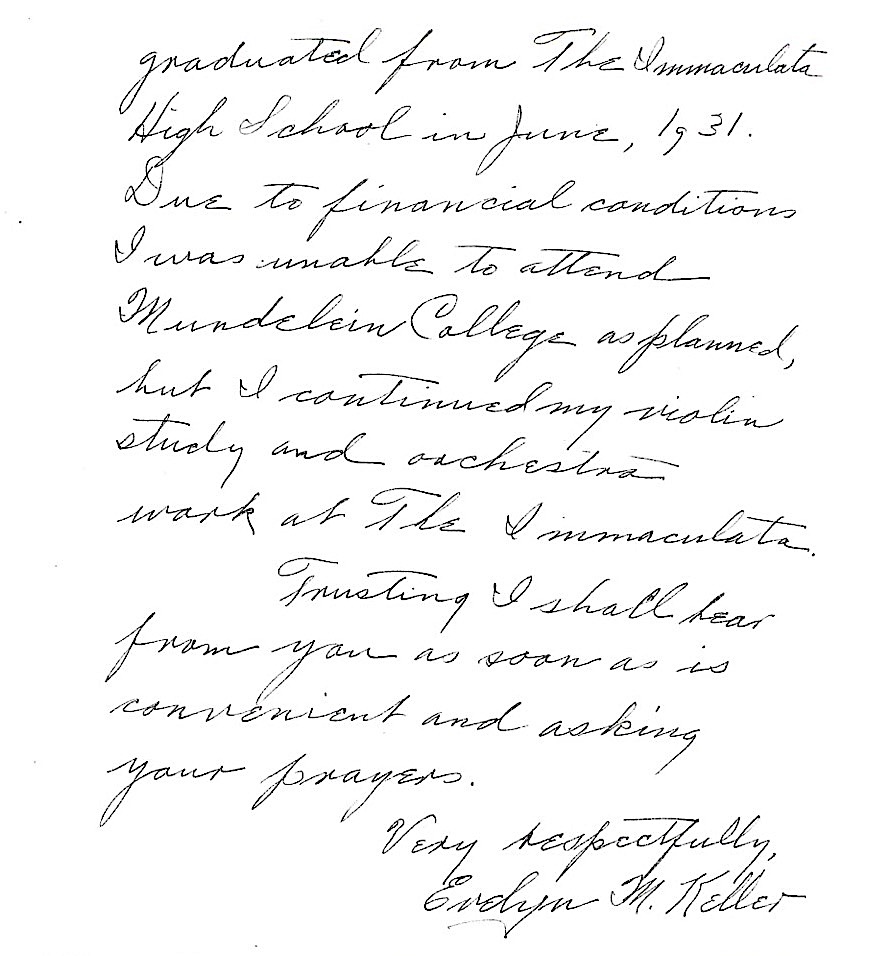}
\\
\end{centering}
\caption{1932 letter from \ken\  explaining that her entrance into the convent was delayed because of her family's financial circumstances.}
\label{fig_KellerLetter}
\end{figure*}

Mary Jerellen Tangney, BVM, wrote in 1932, ``Evelyn's high school teachers, without exception, agree that she is an excellent student and highly intellectual. From the comments of Evelyn's former teachers I gather that her ability is greater in English. I understand that she has done expert work in journalism and possesses unusual possibilities in that field."\cite{jerellen} She noted in contrast that although Evelyn's ``technique" in violin and French horn were good, her intonation was poor.
Evelyn's school work in her last two years of high school suffered ``due to illness in the family and additional home responsibility. ... 
For more than a year the Keller family has been suffering severely from the depression. Mr. Keller is still employed but his present salary covers little more than his carfare. In hopes of the financial difficulty clearing, Evelyn put off entering [the BVM congregation]  last September. Now the family is in a worse way than before and unable to see any relief in the future."\cite{jerellen} The Immaculata sisters agreed to help with her expenses, and she
entered the convent September 8, 1932, at age 18. (See {\bf Figure \ref{fig_mtcarmel}} and {\bf Figure \ref{fig_KellerLetter}}). 
She was formally received into the community on March 19, 1933, and it was then that her name changed from Evelyn to Kenneth.

For the next three years, \ken\ focused on spiritual development and also attended Clarke College in Dubuque and Mundelein College in Chicago. Both institutions were women's colleges founded by the BVMs, and \ken\ completed her basic education and practice teaching there.  After taking her first religious vows on March 19, 1935, she began her teaching career, assigned for the next 29 years to work in elementary and high schools in Illinois and Iowa ({\bf Table \ref{tab:table1}}). 
On August 15, 1940, \ken\ took her final vows, permanently committing herself to the BVM congregation.
In those times, sisters typically finished their academic degrees in summer school so as not to interfere with their teaching assignments; they joked that they received their education ``on the installment plan." \ken\ accumulated additional college credits from Clarke and Mundelein, as well as  13 credit hours from  DePaul University, Chicago, which awarded her B.A. in Mathematical Sciences, with a minor in Latin, in 1943. This  allowed \ken\ to move from elementary to high school teaching (See {\bf Table \ref{tab:table1}}). While teaching, she took 30 hours of course work at DePaul between 1946 and 1952  and earned an M.S. in Mathematics \cite{resume}.

\ken\ was not finished with her studies, however. She took an education class at Edgewood College, Madison, WI, in 1952; zoology from Loyola University, Chicago, in 1953; a workshop in radioisotopes from Iowa State Teachers College in 1955; three more education courses from DePaul in 1956; and 8 credit hours of mathematics from Purdue University (General Electric Fellowship), West Lafayette, IN, in 1957.

\section{INTRODUCTION TO COMPUTING}
\ken's life took an interesting turn when, as a high school math teacher on the west side of Chicago in her mid-40s, she ``read the signs of the times and as early as 1961 \cite{resume} responded by enrolling at Dartmouth College in Hanover, New Hampshire for her first workshop in computer education." \cite{Eulogy} 
As \ken\ told it, ``I just went out to look at a computer one day, and I never came back. ... It looked to me as if the computer would be the most revolutionary tool for doing math that I could get." \cite{witn75} 

Thomas Kurtz, Dartmouth Professor of Mathematics, says that sometime between 1959 and 1961, ``with the assistance of the National Science Foundation,
I began a summer program to introduce high school teachers to computing,
with the LGP-30 as the computer.  These teachers first learned how to use
the machine, and then were required to write simple (?) programs during the
eight to ten weeks of the program.  Sister Mary Kenneth Keller was one of 
the participants." \cite{Kurtz20190111} 

There were no female undergraduates  at Dartmouth at the time, although some graduate programs had a small number of women. The presence of women in the computer center was a bit unusual, and the presence of a Catholic sister covered from head to foot by a black habit with a bit of white starched material holding her veil would certainly have been a unique sight. Kurtz writes,
``I can recall having to find
      a rooming place in town for Sister Mary Kenneth.  (I don't recall whether the male
     participants in my summer program stayed in a dorm, or in town. although
     I don't recall having to find rooms for anyone other tha[n] Sister Mary Kenneth.)" \cite{Kurtz20190111}
     
Dartmouth obtained their desk-sized LGP-30 computer in 1959. Anthony Knapp, who worked on the summer program as a Dartmouth undergraduate student, recalls that programming was done in assembly language. ``There was a sign-up sheet, and one simply had the computer all to oneself for an interval; then it was somebody else's turn.  It had a rotating drum for memory containing 4096 32-bit words.  In each word of memory the right-hand bit was always zero, but it could be one in the accumulator and on input.  The left-hand bit was a sign bit.  The machine language used one word for each instruction, four or six bits for the instruction code and eight bits for the associated address.  The arithmetic was fixed-point arithmetic; some of the arithmetic instructions assumed the decimal point was at the left, and some assumed it was at the right.  The assembly language was very little different; the instruction codes were replaced by letters, and an address would be given by two integers from 0 to 63, track and sector." \cite{Knapp} Before that, boxes of punched cards were carried to MIT to run on their computer. The participants in the summer program learned to use the computer to complete simple programming tasks, such as playing tic-tac-toe. Contrary to published reports (e.g., \cite{Gurer}) Kurtz wrote that  \ken\ ``had absolutely nothing to do with the creation of the BASIC computer language," \cite{Kurtz20190111};  the language was first released in May 1964.

\begin{table}
\begin{centering}
\begin{tabular}{| l l l l| } \hline
1935-36 &
St Callistus &
Chicago, IL &
Teacher: 7th and 8th grade \\

1936-36 &
Our Lady of Victory &
Waterloo, IA &
Teacher: high school          \\

1936-38 &
Holy Name &
Marcus, IA &
Teacher: 7th and 8th grade \\

1938-43 &
St Charles &
Chicago, IL &
Teacher: 7th grade boys \\

1943-51 &
Our Lady of Victory &
Waterloo, IA &
Teacher: high school Math/Physics \\

1951-55 &
Sacred Heart &
Fort Dodge, IA&
Principal \\

1955-61 &
St Mary HS &
Chicago, IL &
Teacher: high school Math \\

1961-63 &
Heelan HS &
Sioux City, IA &
Teacher: high school Math \\

1963-64  &
St Joseph/Alleman &
Rock Island, IL &
Teacher: high school \\

1964-65  &
University of Wl  &
Madison, Wl  &
Graduate student \\

1965-73  &
Clarke College &
Dubuque, IA &
College Teacher \\

1973-83  &
Clarke College &
Dubuque, IA &
College Administrator\\ \hline

\end{tabular}\\
\hspace{2pt}\\
\end{centering}
\caption{\ken's career assignments \cite{resume}}.
\label{tab:table1}
\end{table}

\section{PhD STUDY}

Thomas Kurtz recalls that \ken\ ``had connections with a small college in her area." \cite{Kurtz20190111} This probably refers to Clarke College, a women's college in Dubuque, Iowa, founded in 1843 as St. Mary's Female Academy, later called Mount St. Joseph, and renamed for the founder of the BVM congregation, Mary Frances Clarke, in 1928.
At some point, Mary Benedict Phelan, BVM, president of Clarke College, decided that students there should be prepared for the information age and that they needed a sister to head up the effort. \ken\ was sent to the University of Wisconsin, Madison, to complete a PhD degree and prepare for this work \cite{mone,bone}. Although she lists her enrollment as 1962-1965 \cite{resume}, her break in teaching assignments was only one year, 1964-65, the year she completed her degree (see Table \ref{tab:table1}), so again much of her study must have been during summers. She completed 15 credit hours in computer science and 21 credit hours of research at the rate of 6 hours per year until her final year \cite{resume}. \ctwo\ was a student at Wisconsin at the same time and remembers that both of them would work until midnight and then go to an all-night drugstore for ice cream sundaes. She jokes that because of this, \ken\ had to teach her how to let the pleats of her habit out.

\ken\ was supported by a National Science Foundation Fellowship (1962-64) and then by a university fellowship. Her academic advisor at Wisconsin was Preston Hammer, Professor of Numerical Analysis and (as of 1964) Professor of Computer Science. His colleague Seymour Parter recalls that  ``Preston was a remarkable man, absolutely a remarkable man.... He went into research that was closely related to computing," different from anyone else at the university \cite[00:23:27]{Parter}. Hammer worked in topology, logic, and numerical integration. He broke away from the Mathematics Department to begin a Department of Numerical Analysis in 1961, perhaps because he believed that ``the schism between pure and applied mathematics ... is accentuated by great fools," \cite{HammerEd} and he seems to have been a man who did not suffer fools easily.
 Parter recalls \ken\ ``barely":  ``She was an interesting woman, she learned numerical analysis....
I came in '63 so I guess I must have seen her around. She never took a course with me. I think she was busy writing her thesis." \cite[00:58:03]{Parter}

\ken\ begins her dissertation \cite{phd1}, ``Inductive Inference on Computer Generated Patterns," with the declaration, ``The basic problem of this thesis is the exploration of an approach to the mechanization of inductive inference."  
The idea is to automatically generate the $n$th mathematical expression in a series, given the first $k$ expressions. 
As validation, \ken\ implemented a FORTRAN program to determine the $n$th derivative of a function, given the first $k$ derivatives. She remarked that list processing languages such as LISP would have been more convenient but were not available to her.
She referred to a 1962 publication describing a program for automatic differentiation \cite{hanson}, but this, of course, was not her goal. She wanted to demonstrate that algorithms could perform tasks like differentiation through learning-by-example, rather than by a rule-based process. Although this approach was out of favor for many years, it has recently had a renaissance in so-called ``deep learning models" in artificial intelligence and now dominates the field.

She implemented a module to compute derivatives (used to generate input to her program and check results), a parser for FORTRAN-style arithmetic expressions, modules to test for common patterns in coefficients of the $k$ expressions (e.g., arithmetic series), a module to generate the hypothesized $n$th derivative, and a module to generate a FORTRAN expression from the internal representation of the derivative. From her description, it seems that she could handle polynomials and expressions involving standard FORTRAN functions (trig functions, log, exp, etc.) as long as the example derivatives were written in a consistent way. For example, if the $j$th derivative is $(x-2)^3$, then the $(j+1)$st derivative should be written as $3(x-2)^2$, not $3x^2 -12x +12$.

\ken\ did not claim that her ``results obtained in the domain of analytic differentiation were important in themselves," but that they demonstrated that her programs ``do mechanize an inductive inference. They effect a `jump' from a kth observation to an nth observation, and in problems of the type tested in this experiment, this is accomplished in less than 30 seconds per problem." (pp. 46-47)

She concludes with her first statement on the relation between human and artificial intelligence: ``The claim that this parallels the activity of the mathematician who generalizes on a symbol manipulation routine after some number of repetitions must, of course, be qualified." (p. 46)

\begin{figure*}
\begin{centering}
\includegraphics[width=.99\columnwidth] {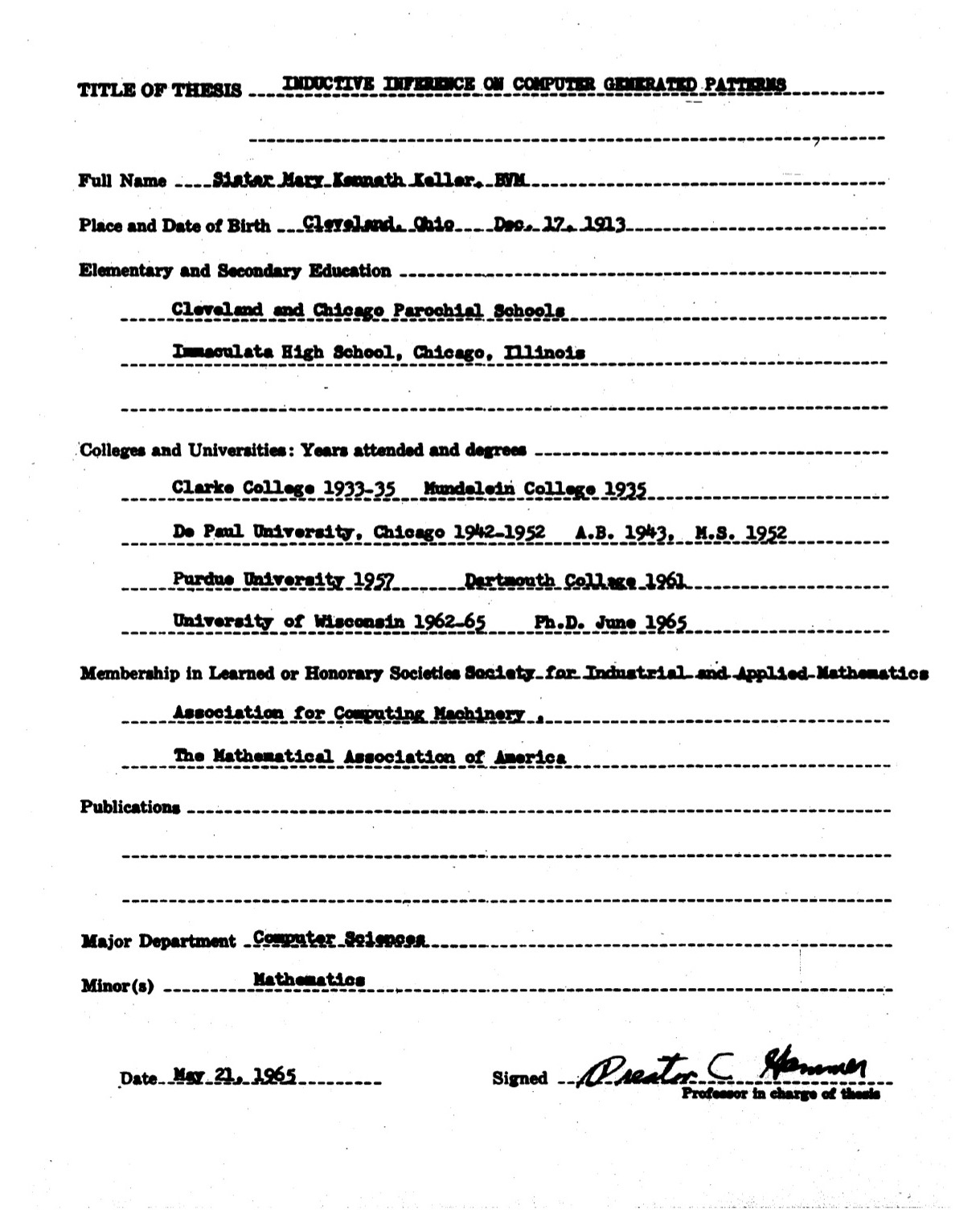}\\
\end{centering}
\caption{Page 67 of \ken's thesis \cite{phd1}.}
\label{fig_thesis}
\end{figure*}

The thesis bears the signature of her advisor Preston C.~Hammer (May 21, 1965), as well as that of Ralph L.~London, who worked in programming languages and verification, and Eldo Koenig, an engineer with interests in man-machine interactions and  robotics. The ``major department" is listed as ``Computer Sciences" and the minor as ``Mathematics" \cite[p67]{phd1}; see {\bf Figure \ref{fig_thesis}}. She listed membership in the Society for Industrial and Applied Mathematics (SIAM), the Association for Computer Machinery (ACM), and the Mathematical Association of America (MAA).

After 29 years of K-12 teaching and 33 years of college-level studies, \ken\ had finally completed her formal education. She was 51 years old.

\section{WORK AT CLARKE COLLEGE}

\ken\ arrived at Clarke College with a vision: ``Automation, or more precisely, cybernation, is a fact of our lives. Its impact is swift and in many ways silent. Our sense of responsibilty should make us wish to be informed and to inform our students. ... Furthermore, the computer can give a new dimension to the education we offer." \cite{vista66} {See {\bf Figure \ref{fig_circuit}}.

\begin{figure}
\begin{centering}
\includegraphics[width=.6\columnwidth] {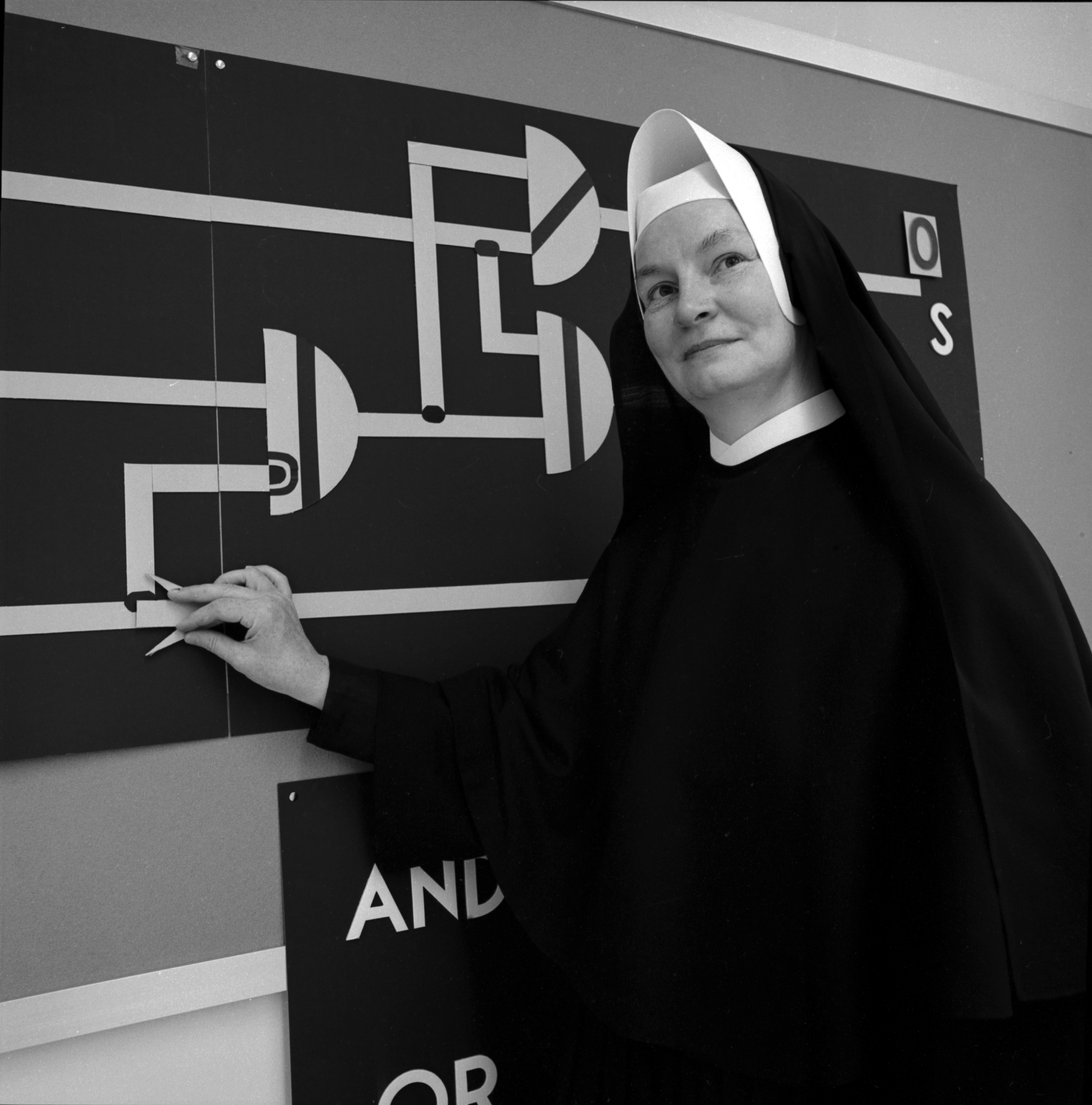}\\
\end{centering} 
\caption{\ken\ in 1965 with a circuit diagram. Photo: Reprinted with permission of the \emph{Dubuque Telegraph Herald}.}
\label{fig_circuit}
\end{figure}

\subsection{Managing the Computer Center}
In a 1964 news release, Mary Consolatrice Wright, Mother General of the BVM Congregation, announced a ``Plan for Center for Computer Sciences" at Clarke College \cite{c64a}, with \ken\ (who had previously taught mathematics courses at Clarke in the summer) to be appointed director in 1965. She asked for prayers for the success of a proposal to the US government to cover half the cost of computer installation. Around the same time, Mary Benedict Phelan, BVM, Clarke College President \cite{c64b}, noted that ``a computer for educational purposes" was to be installed in summer, 1965.``Courses will serve undergrads, teachers, and industrial personnel: intro programming, cs for teachers, cs for research. ... It is our wish that the computer be available to all area colleges and universities for instructional purposes." In addition, ``Sister Mary Benedict noted that many educators believe that, within a few years, access to a computer will be as necessary for a college faculty as access to a library is today." 
The machine was an IBM 1130 with 8K memory, actually installed with its paper tape reader in 1966 and  upgraded to punched card input in 1969 \cite{mem}.

\begin{figure}
\begin{centering}
\includegraphics[width=.99\columnwidth] {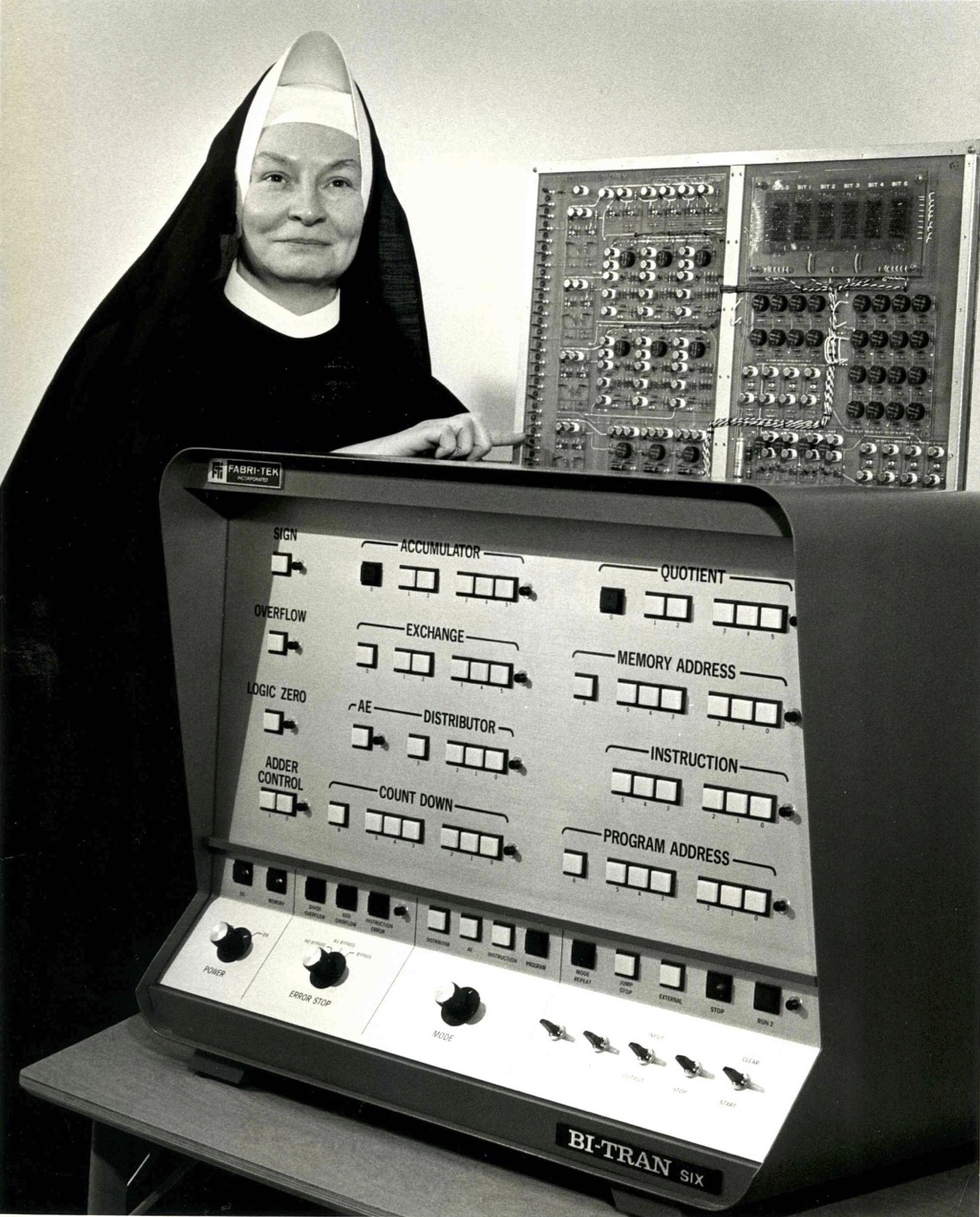}\\
\end{centering}
\caption{\ken\ with Bi-Tran Six computer. Photo credit: Clarke University Archives.}
\label{fig_bitran}
\end{figure}

Before the arrival of the IBM 1130 \cite{vista66}, the computing power consisted of a desktop computer, the Bi-Tran Six, shown in {\bf Figure \ref{fig_bitran}}, with 128 6-bit words of memory, visible circuit boards, and programming in assembly language \cite{bitran}. 
Keeping the computer facility up-to-date on a shoe-string budget was a continuing struggle, and  Clarke participated in a 1967 proposal by the University of Iowa and ten other participating schools for a Remote Computer Center (RCC) to be located in Iowa City. Clarke's contribution was space for a terminal, \$10,000 in salary funds, supplies, and overhead \cite{c67b}. The proposal was successful, and \ken\ continued to participate \cite{exposing}, but she later had misgivings about the level of user support and documentation provided by RCC \cite{c78d}.

\begin{figure}
\begin{centering}
\includegraphics[width=.99\columnwidth] {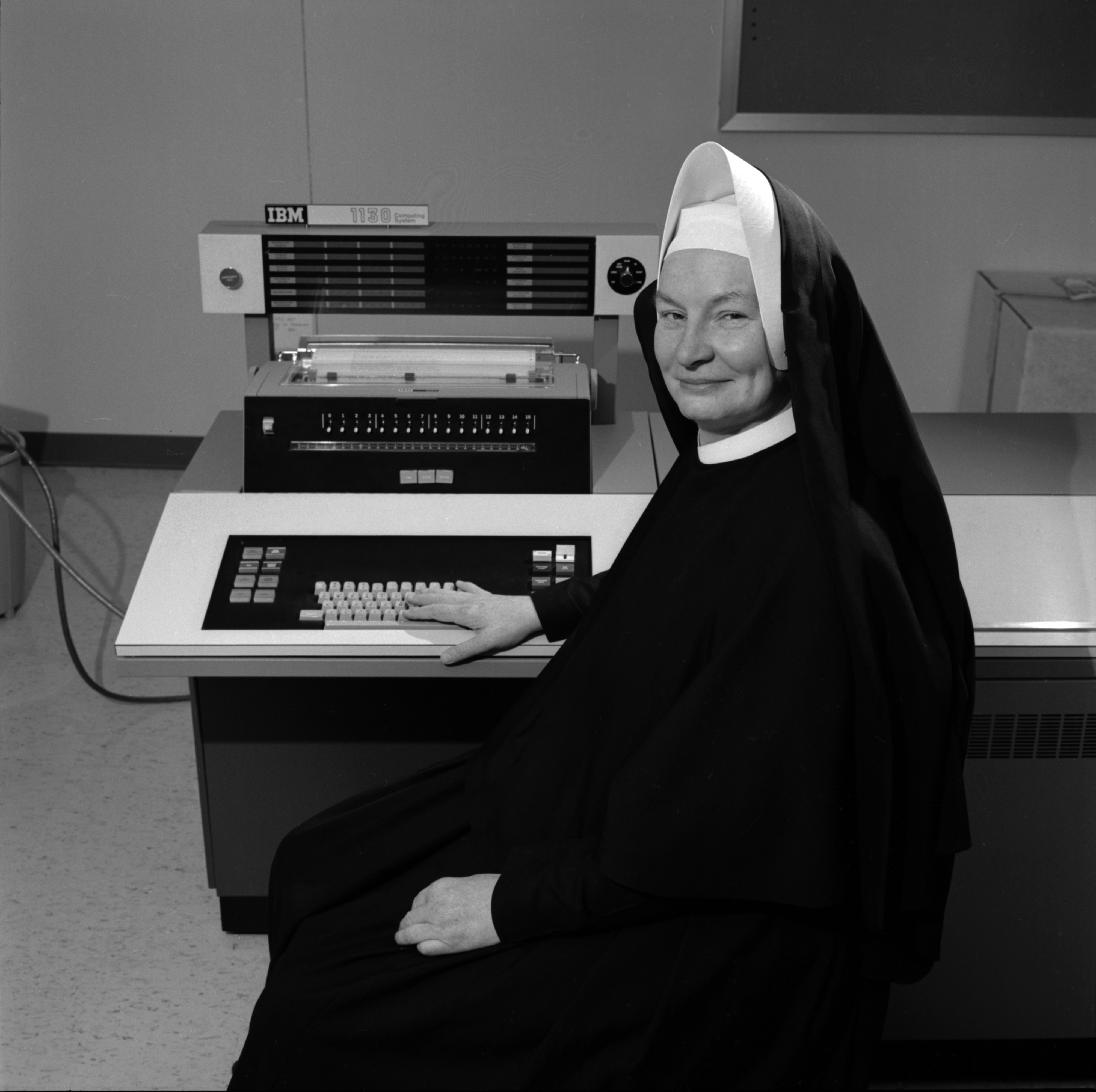}\\
\end{centering}
\caption{\ken\ in September, 1966 with Clarke's IBM 1130. Photo: Reprinted with permission of the \emph{Dubuque Telegraph Herald}.}
\label{fig_1130}
\end{figure}

In 1977, \ken\ retired Clarke's IBM 1130, shown in {\bf Figure \ref{fig_1130}}, replacing it with a vintage 1968 IBM 360-40 with 256K memory, a gift of a Mr. Wahlert of Dubuque Packing Company \cite{mem,c77d}. \ken\ noted that the machine was capable of servicing 31 interactive terminals and would be used for all of Clarke's ``academic and administrative applications." \cite{c78a,c77c} Because of its greater speed, 81\% of the computer time could be devoted to academic use \cite{c78a}.

\ken\ worked long hours and lived on campus in a small bedroom on the top floor of Mary Bertrand Hall. She also took time to participate in college governance, chairing the Clarke College Forum in the early-1970s, shaping college policy \cite{fink}.

To serve administrative computing, Clarke's computer center developed software systems.  For example, they teamed with Furman University (Greenville, SC) to develop a database system, written in FORTRAN IV,  called the RELCV Information System, for the 8K IBM 1130, eventually adapted for the IBM 360 and other machines. It could sort and select fixed-length records from a disk file, generate and print reports, and perform statistical analysis. It was used to process student admissions at both schools. By October 1971, 250 schools had acquired it at a cost of \$25 for a deck of cards, systems manual, and user manual \cite{c71}. Clarke also employed a systems analyst who designed systems including a gifts and grants program, payroll, and accounts payable, used by Clarke as well as the University of Dubuque \cite{cundat}.

\begin{figure}
\begin{centering}
\includegraphics[width=.99\columnwidth] {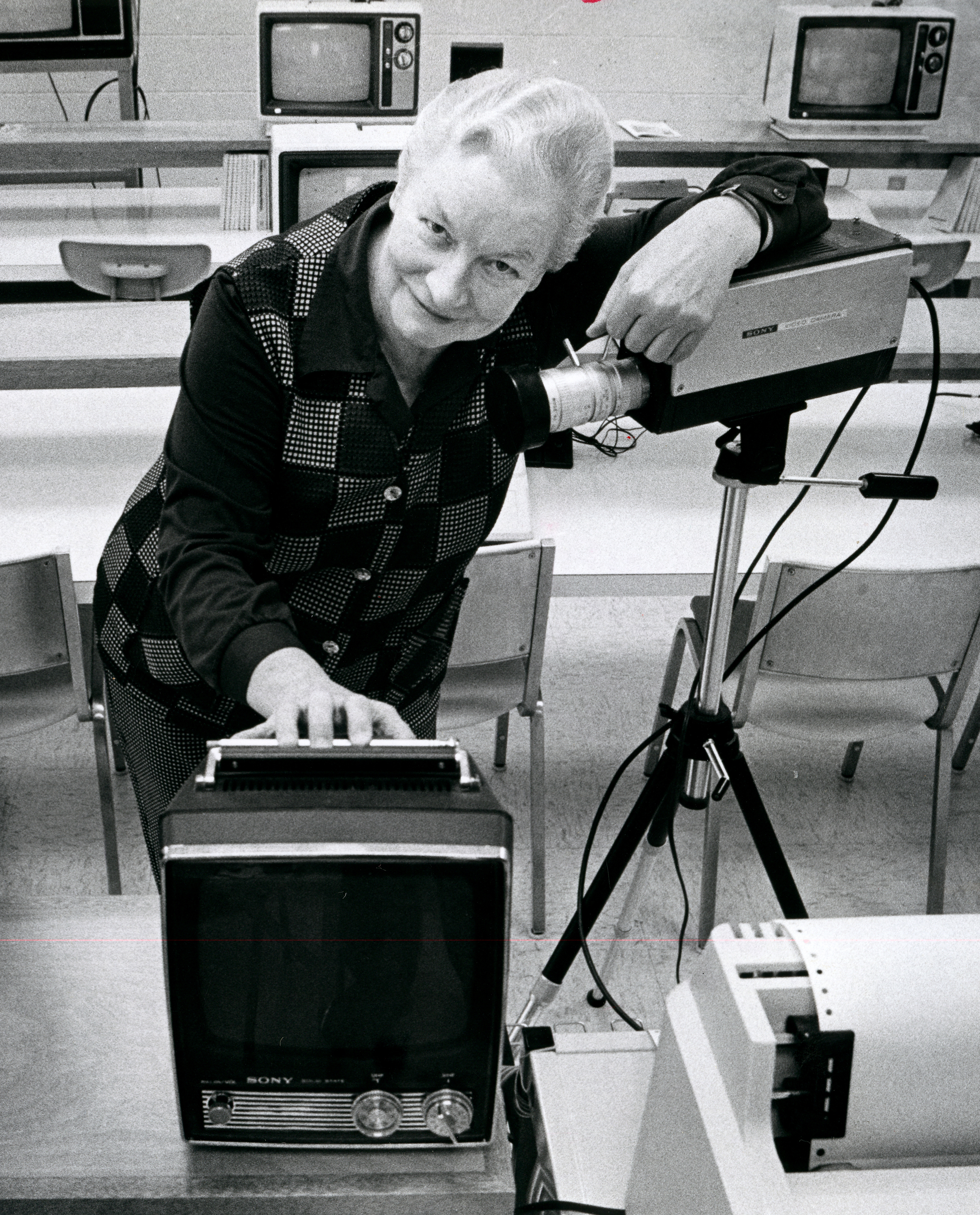}\\
\end{centering}
\caption{\ken\ in Clarke's Computer Center in September, 1980. Photo: Reprinted with permission of the \emph{Dubuque Telegraph Herald}.}
\label{fig_micro}
\end{figure}

In addition to managing academic and administrative computing at Clarke, \ken\ kept her hand in research; in 1975, she was collaborating with  Raymond Martin at Wartburg Theological Seminary in Dubuque, using the computer to compare the Hebrew, Aramaic, and Greek texts of two books of the Bible \cite{witn75}, research that was related to Martin's previous work in \cite{martin}.

By 1978, \ken\ advocated using microprocessing technology \cite{c78a} in place of large, central computing systems. In 1980, she received a \$206,676 NSF grant entitled ``Comprehensive Application of Computer Technology to Science Education," \cite{c80a} to equip a lab with 20 microcomputers, shown in {\bf Figure \ref{fig_micro}}, for classroom use. The machines could be configured with color graphics, sound (music and speech synthesis), light pens, plotters, and more, one more step toward making the machines useful to liberal arts students. The newly-named Keller Computer Center \cite{csb} had tripled in size from its original classroom \cite{vista66} to 2,750 square feet \cite{tele80} reclaimed from Clarke's laundry building, and was also equipped with a IBM 4331 with a megabyte of memory \cite{bvm03}.
By 1982 there were 21 Apple computers on campus: 16 in the computer lab and one each in the library, the Instructional Resource Center, and the Departments of Music, Chemistry, and Business.

Under \ken, the computer center was a family-friendly environment. After Kathy Decker (Clarke class of 1974) was hired, despite the cramped quarters, \ken\ provided nursing and play space for her children \cite{onc}. Kathy Decker eventually succeeded her as Director of Administrative and Academic Computing in 1983 \cite{onc}, upon \ken's recommendation \cite{c83a}.

\subsection{Contributions to Clarke College education}
The  December 1964 ``Faculty Focus" included an article by Sisters Martha (Briant) Ryder and Harriet (Agneda) Hollis, BVM faculty members at Clarke College,  that quoted John Kemeny, a professor whom \ken\ had met at Dartmouth: ``In addition to the small number of people who will personally use [computers], millions will be indirectly affected by their very existence. It is essential that a student receiving liberal arts education should, in the future, be acquainted with the potential and limitations of high speed computers." \cite{c64c} 
Guided by this philosophy, a natural consequence of BVM foundress Mary Frances Clarke's directive to ``keep our schools progressive with the times in which we live" \cite{maryc}, \ken\ became an evangelist for computer science, determined that the women graduating from Clarke would be ready for the computer age.
Until 1968, she was ``a one-woman computer sciences department." \cite{tele65,mem}
Immediately upon taking her position at Clarke in Fall 1965 \cite{c65x}, \ken\  initiated a 3-credit-hour introductory course and established a minor concentration in computer science, making Clarke the first private college in Iowa to do this. She initially did not want a computer science major, believing that the liberal arts should be encouraged \cite{mone}.

``Sister Computer" \cite{sone} noted in 1965 that a broad mathematical background was not necessary to study computing; ``There is one requirement for computer work, however: confidence. The computer doesn't make mistakes and it is very patient with you while you correct your mistakes, but you need confidence in your own ability to de-bug the program so that it will run." \cite{c65x}
A year later, she had identified two other essential qualities: humility and patience. ``The computer doesn't make mistakes, and it's hard on one's ego to have to assume all the blame every time something goes wrong. That's where humility comes in. And then you get to practice patience by correcting your mistakes." \cite{vista66}
Sheila Blaha Sullivan (Clarke class of 1970) recalled, ``We learned that we were capable of handling problems if we would just poke at them long enough from another direction." \cite{onc}

\ken\ was a gifted teacher and communicator. In 1967, R. (Richard) Buckminster Fuller (designer of the geodesic dome, called  ``Genius of our Time") accepted the Thanksgiving Day Award at Clarke. He was so impressed with \ken\ that he asked to return to campus in September 1968 (see {\bf Figure \ref{fig_buckminster}}) for two days of intensive consultation with her on how computers could augment his work \cite{th68}. 
Fuller said, ``I wondered if I'd be able to do this successfully at the age of 73," but he found that after \ken's tutoring, he could, ``slowly." \cite{mr68}
``He believed that technology should be directed to `livingry' rather than to military power and weaponry," \cite{onc}, a view shared by \ken. 

\begin{figure}
\begin{centering}
\includegraphics[width=.99\columnwidth] {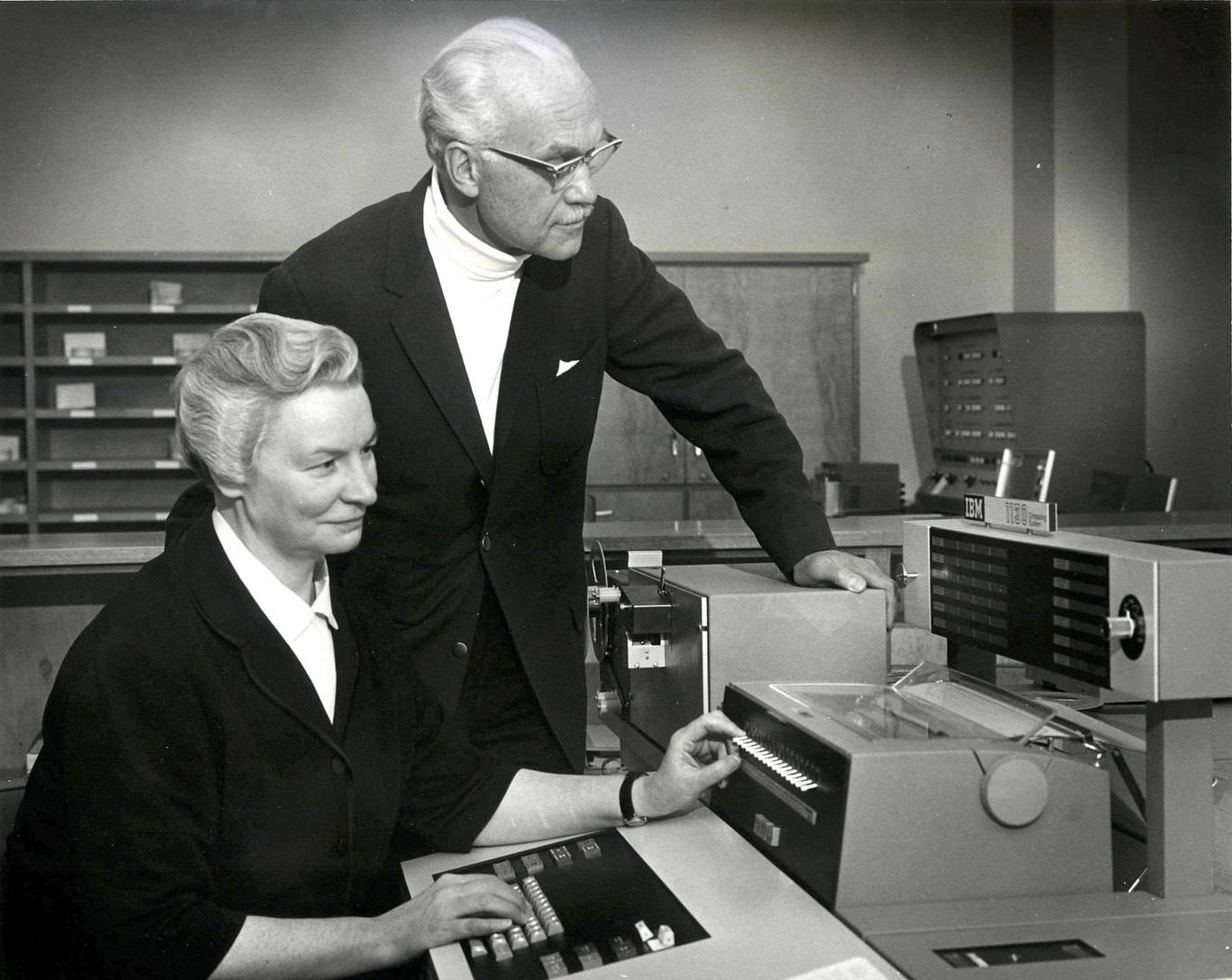}\\
\end{centering}
\caption{``73-year-young R. Buckminster Fuller is taking a two-day crash course in computer science and programming from Sister Mary Kenneth." \cite{th68}. Photo credit: Clarke University Archives.}
\label{fig_buckminster}
\end{figure}

\ken\ was a witty teacher, engaging each person as she walked around the room \cite{done}. In addition to regular courses, she taught evening adult education in FORTRAN and assembly language \cite{mone}.
She loved art and drama and  ``coerced" her academic colleagues into learning computing, telling artists and musicians, ``You have to know this!" \cite{cone} She got them ``hooked up on that Apple computer,"  and even told IBM representatives that they needed to build a machine like that, rather than just producing large mainframe computers \cite{bone}.
She was ``just so pleased if people used the computer in a new way." \cite{bone}

Although she was a demanding educator, Sister Kenneth used humor to facilitate learning and to make her students feel empowered. At one point, using (now archaic) terminology that specified relations among computer processes, she programmed the computer to accompany every error message with the declaration, ``You are the master, I am the slave." \cite{res93}.

By Summer 1971, \ken\ had placed Clarke senior Sue Park in a work/study program at Mercy Hospital, working on a computer-assisted food management system. The news announcement noted that \ken\ ``worked on the development of such a program this summer at Tulane University." \cite{c71a}

In 1973, \ken\ transitioned out of teaching into full-time administration.
In 1977, Clarke merged the Department of Economics / Management Science into the Department of Computer Science to form the Department of Computer and Management Sciences. \ken\ justified this by noting that most Computer Science graduates from Clarke were going into business and industry, and she wanted to form a Women in Management Program to integrate the disciplines \cite{c77e}. This doubled the size of the faculty, from two to four. By 1980, Marianne Joy, BVM  had succeeded \ken\ as department chair \cite{c80a}.

Clarke established a Bachelor of Arts degree in 1979, with 7 graduates in 1980 \cite{mem}, all of whom had jobs as programmers or analysts, mostly with local firms \cite{c80a}. Many sophomores and juniors had summer or work/study jobs \cite{c80a}. A Bachelor of Science degree was added in 1982 \cite{mem}.
By that time, Computer Science at Clarke had 51 students, including 30 masters candidates \cite{c82a}.

\ken\ continued her advocacy for computer education in secondary schools, running NSF Summer Institutes in Mathematics for Secondary School Teachers from 1968 to 1972 \cite{c71b,resume}.
In 1981, Clarke established a summer Masters Program in Computer Applications in Education \cite{c8212}, numbering ``among a very few such programs nationwide." \cite{c82a} 
Fifty students made up the first cohort, completing 9-12 credit hours in education, 15-18 in computer science, and 3-6 in electives.
David Fyten, a Clarke faculty member, was quoted as saying, ``Many teachers feel that the introduction of microcomputer education has rejuvenated their desire to stay in teaching." 

In the January 1983 semester, Computer Science ran 10 courses with 19 sections averaging 25 students per section, taught by 5 FTE instructors \cite{c83a}.

In 1983, \ken\ was asked to prepare a 5-year plan for the computer center. She complied but noted that ``A five year plan for the computer center can only be a projection from the present, and not too real. Consider that five years ago almost nothing that is in the center today existed. The field is in even greater flux today." \cite{c83a} She did know one thing with certainty: ``A search for a new director of the computer center should be started," since she was dealing with breast cancer after surgery in August, 1982 \cite{eug} and knew that she needed to retire, at age 69.

\section{COMMUNITY SERVICE}

\subsection{Service to industry}
During the late 1960s, \ken\ established herself as a local expert on computing hardware, software, and applications; see {\bf Figure \ref{fig_hospitalconf}}. She was much in demand by local government \cite{bvm03} and local companies seeking advice on their hardware and software needs, through one-on-one consulting work as well as forums such as a Clarke ``bank computer seminar" to serve the local financial community \cite{c67a}. She directed seven Iowa Technical Service Seminars for Business and Industry between 1967 and 1970 \cite{resume}. She consulted for the City of Dubuque (1968-1970, 1978-1979) and Mercy Hospital (1966-1967, 1970, 1978) \cite{resume}. A wide variety of industries sought her advice, from Ertl Toy Company to Magma Copper Mines to Cullen Kilby Caralan Engineering \cite{resume}. ``People called her up, and she solved their problems." \cite{mone} She also spent summers in 1969-1971 consulting in Illinois \cite{witn75,resume} for the State Office of Budget Management \cite{res93} developing ``a state-wide management information system" \cite{ccr71}. 

The relationships she developed supported Clarke in the 1970s, for example, through the donation of the IBM 360-40 and through contracts, such as one for Clarke to develop software for John Deere Dubuque Works \cite{c79a}.
``Almost single handed she equipped the computer lab by giving lectures in colleges and universities, conferences, conventions, lunches, dinners, wherever she could reach an audience and be paid to do it." \cite{bvm03}
In those lectures, \ken\ often put considerable emphasis on ethics, saying that computers should be used to open up new possibilities in industry rather than replacing current workers, and that computers should be used to alleviate poverty and ignorance \cite{st66}.

\begin{figure}
\begin{centering}
\includegraphics[width=.99\columnwidth] {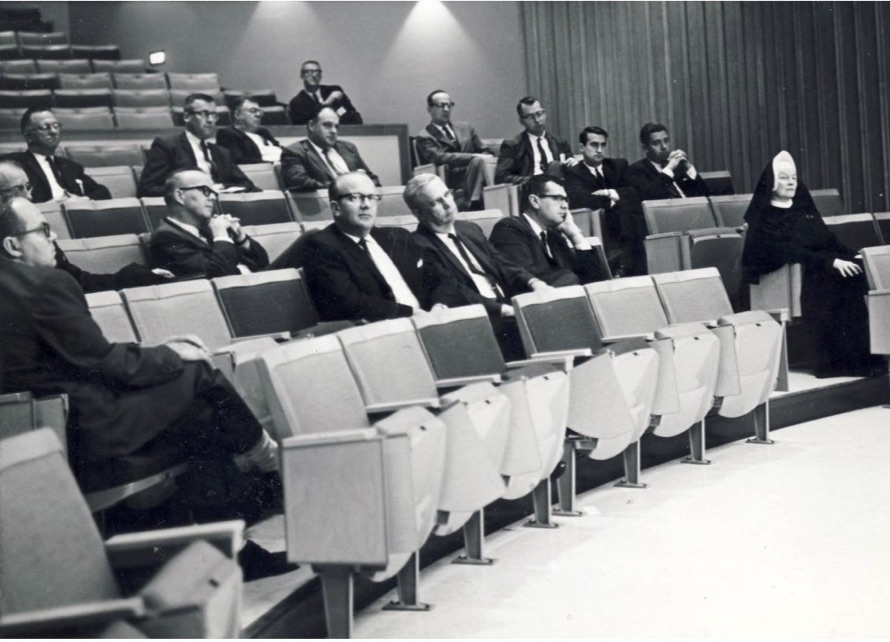}\\
\end{centering}
\caption{\ken\ at an August 2-3, 1967 conference at Clarke College on Hospital and Medical Applications of Computers.  Photo credit: Clarke University Archives.}
\label{fig_hospitalconf}
\end{figure}

In a 1967 speech on ``Computer Applications from a Management Point of View," \ken\ noted that, ``By 1965 third generation hardware was being delivered which, in theory, at least, should have greatly extended the use of computers to the areas of control and decision-making." She  noted that these advances were slowed by a shortage in computer programmers and the overhead of rethinking the boundaries affecting dataflow within companies, leading to ``third generation hardware with second generation programming systems." She predicted  that soon even small organizations would use computers, either by using dedicated machines or purchasing time-sharing. Computers would also become management tools. 
\begin{quote}
The concept of a data base where all items are organized on a tree with inter connected branches would seem to be the correct principle of operation for the future, ... a complete departure from the structuring of data files in the traditional sense where a payroll file belongs to one department and a separate personnel file [exists] for the use of the personnel department. ... There will likely be a greater use of ... visual display output which is meant to be a momentary presentation of information for view by the executive on an on-call demand basis, whereas the hard copy output will be limited to traditional reports and output documents.
\end{quote}
She also highlighted the use of computers in process and production control and in simulation. All of this depended on a sufficient supply of knowledgable programmers, and she worked hard to establish a pipeline of educated candidates from Clarke and other small colleges and universities.

\subsection{Service to education}

\ken\ was a founding member of the College and University Eleven-Thirty Users Group (CUETUG) which, as IBM 1130s were retired, was renamed the Association of Small Computer Users in Education (ASCUE) \cite{cath85,c85,resume}. She spoke at their meetings (e.g., \cite{fieldexp,bus,micro}) and represented ASCUE on the steering committee for the National Educational Computing  Conference. She  served  as a Board member (1974-1976) and as Public Relations Director (1977-1984) until shortly before her death \cite{ascue}.

\ken\ was active in ACM's Curriculum Committee for Undergraduate Computer Science, and the highly influential ACM Curriculum '78, a model for undergraduate computer science education, lists her as a contributor \cite[p.166]{ACM78}.  She also worked on the Masters-level curriculum \cite{cce}.

\ken\ spoke at many professional meetings on computer facilities for small colleges (e.g., \cite{uc}) and computing in undergraduate curricula \cite{curric}, and various colleges and universities called upon her for advice on computer science curricula. 

As her reputation grew, so did the demands on her time, and in April 1980, she testified before Congressional subcommittees on information technology in education \cite{mone,cce}.

The IEEE Computer Society honored Sister Kenneth by establishing the Mary Kenneth Keller Computer 
Science \& Engineering Undergraduate Teaching Award, which has been given most years since 1999 \cite{IEEETeaching}.

\section{TEXTBOOK AND EDUCATIONAL MODULES}

\subsection{Textbook}

\ken\ collaborated on a 1973 textbook entitled {\em Mathematical Logic and Probability with BASIC Programming} \cite{Dorn}, based on an earlier textbook by Dorn and Greenberg entitled {\em Mathematics and Computing with FORTRAN Programming} \cite{DornGreen}. The newer version omited applications in linear algebra and calculus and used the  more accessible BASIC programming language in order to reach a broader audience. 

\ken's contributions began with a 43 page introduction to BASIC programming (Secs 1.1-1.13, 2.4), developing the idea of an algorithm through the solution of a particular system of two linear equations. She  then generalized the method to solve  any two linear equations (including checking for the no-solution condition), wrote a flow chart for the algorithm, and then implemented the algorithm in  BASIC. After that motivational example, she systematically presented input/output statements, constants and variables, assignment statements, {\tt if/then} and {\tt goto}, {\tt stop} and {\tt end}, {\tt for/next}, arrays, functions, and boolean operations. Her attempts to prevent student errors were remarkable; for example, she carefully explained that each array is named by a single letter but the names of  other variables may include a single digit after the letter. The level of clarity is noticeably higher than that of the following section about computer arithmetic, taken from Dorn and Greenberg, which explained the intricacies of rounding, normalization, and loss of precision in base 10; the text noted that although many computers use base 2 or base 16, ``This need not concern us here since BASIC always uses powers of 10," (p.40) which would be true only for integer arithmetic.

\ken's approach to pedagogy is most evident in the book's Appendix A, where she stepped students through their first encounter with using a teletype, from sign-on to sign-off. It seemed important to her to provide students with every piece of information they would need to succeed, avoiding unnecessary discouragement.

Two other sections by \ken\ complemented the discussion of {\tt and-or-not} computer circuits by explaining how truth tables define a boolean function and showed how such functions can be simplified using Karnaugh maps. Students were invited to check their work using BASIC programs.

A final section explained how to use random numbers to run simulations, using a baseball batter's history as an example.

\subsection{Work with UMAP}

Following a 2-week curriculum development meeting in Massachusetts in 1975 \cite{bvm03}, \ken\ wrote several educational modules for the Modules and Monographs in Undergraduate Mathematics and Its Applications Project (UMAP). ``The goal of UMAP is to develop, through a community of users and developers,  a system of instructional modules in undergraduate mathematics and its applications that may be used to supplement existing courses and from which complete courses may eventually be built." \cite[facepage]{umap3}. \ken\ structured her contributions in pairs, one elementary and one more advanced. She included exercises, a sample test, answers, and computer programs. She aimed to make the modules attractive to students by choosing topics of interest, such as food, pictures, prediction of the future, and electricity.

Modules U105 and U109 \cite{umap4} concerned food service management. In the elementary module she introduced a matrix with rows corresponding to ingredients and columns corresponding to menu items, and showed how to compute  the cost of various menu items, given unit cost for ingredients and the amount of ingredients needed for various combinations of menu items. In the advanced module, she invited the student to create a nutrient matrix (rows are nutrients and columns are ingredients) and create a diet that meets specified nutritional requirements. Perhaps \ken\ drew upon a 1966 course she had taken at the University of Michigan, Ann Arbor, in ``Applications of Computers in Hospitals." \cite{resume}

Module U106 \cite{umap3} concerned computer graphics. \ken\ walked students through the process of rotating a set of coordinates in a plane (e.g., endpoints for line segments) through a specified angle by multiplication by a $2 \times 2$ rotation matrix. Then she introduced homogeneous coordinates, 3 coordinates for each point in the plane, a common choice in computer graphics. She invited students to explore how to rotate, translate, and scale the endpoints of line segments, including scaling for perspective, using homogeneous coordinates. As aids, she provided FORTRAN programs for reading points, transforming them, and plotting.
Continuing this theme in Module U110 \cite{umap3}, \ken\ invited students to determine how to rotate, translate, and scale sets of points in 3-dimensions using homogeneous coordinates.   

Modules U107 and U111 \cite{umap2} concerned Markov chains. First, using an example of how performance on one exam predicts a student's performance on the next exam, \ken\ introduced the transition matrix $M$ for a Markov chain and showed that the probability of future events can be calculated by computing powers of this matrix. She had the student investigate the limit of $M^k$ as $k$ increases for an example involving career choices of women. In the advanced module, students explored that limit again, using an example of television viewing habits, to demonstrate that the stationary vector of the chain (the long-term probabilities of the model) are found by solving $p M = p$. Using an example involving loan repayment prediction, which has absorbing states, students computed the ``fundamental matrix," which gives predictions of the number of times any state is visited. All of these computations were supported by FORTRAN programs provided in the module.

Module U108 \cite{umap1} presented Kirchoff's laws and Ohm's law and showed how to define a system of linear equations to analyze a (linear) electrical circuit. FORTRAN software was provided for reading matrices, solving linear systems, and writing results. Matrices were tested for singularity by reduction to echelon form, using a tolerance of $10^{-4}$ to define zero. In Module U112 \cite{umap1}, \ken\ invited students to investigate the use of the software on more complicated circuits.

\section{SISTER KENNETH'S PHILOSOPHY OF COMPUTING}

\ken\ began an unfinished book entitled, ``The Computer: A Humanistic Approach." She viewed the computer as an ``idea processor" for text and programs that should be used in schools as early as the fourth grade \cite{idea}. The book was to be a source of elementary, intermediate, and advanced problems amenable to computer solution. She emphasized ``pictorial and graphical representations for solutions" and continued her theme that problems should be chosen to be of particular interest to students.

\ken\ was a visionary.
Her colleagues were in awe of her, astounded by  prescient predictions that wristwatches would become computers \cite{ctwo} and that computers would transform the arts and humanities. ``She'd blow your mind, and that is what she wanted to do." \cite{cthree}
As early as 1966, she envisioned the use of computers in every academic discipline:
\begin{quote} 
We really don't know how people learn. For the first time, we can now mechanically simulate the cognitive process. We can make studies in artificial intelligence. Beyond that, this mechanism can be used to assist humans to learn. It reacts patiently, persistently. It can store the number of paths a student may take, and point out the successes. As we are going to have more mature students, in greater numbers, as time goes on, this type of teaching will probably be increasingly important. In the modern linguistic field, for instance, the whole science of language and grammar may be studied by this method. \cite{midw66}
\end{quote}
She was also aware  of the fears and hopes associated with computing: fears that jobs would be eliminated and automated decisions would be unquestioned, hopes of more meaningful work and ``a super-planner which can aid the world's growing population in production, protection, and decision. It is reasonable to stand with those who hope, but it must be with a sense of responsibility," since ``man has not always used his inventions well." \cite{bvm1} 

Perhaps her most complete exposition of her philosophy was presented in an undated address to the Cedar Rapids Conference on ``The Role of the Computer in the High School."
\ken\ noted, 
\begin{quote}
The computer is an electronic realization based on consideration of modes of thought, strategy of decision making, methods of experimentation, the theory of communication, and in fact, all of the ramifications which belong to the intellectual activities of man.
\end{quote}
She pointed out its potential as a teaching machine, librarian for information retrieval, ``super-clerk" for reducing the monotony of administrative tasks, and a linguist for foreign language translation.
\begin{quote}
Of course it is a mathematician, but please don't think of it as just a special adding machine or even a super slide rule. It is just a historical accident that such connotation [is] attached. It is essentially a symbol manipulator.
\end{quote}
She envisioned the computer improving education and freeing teachers to have ``more human contact with your students and fellow teachers."
\ken\ gave an emphatic ``yes" in answer to the question of whether computation should be taught in the high school: 
\begin{quote}
Every citizen has a right and a duty to have a knowledge, commensurate with his capacity, of the important forces and instruments which shape his civilization. Not only, as I have said, would we be presenting a distorted, truncated view of subject areas [if computation were omitted], but we would be failing to capitalize on the ease of introducing basic concepts to young adults as compared to mature adults, and we would be denying many that knowledge altogether.
\end{quote}
\ken\ believed that every high school student should be able to program a computer and to understand the internal binary logic systems employed by the machines. It is worth noting that in the US even now, more than 40 years later, this goal has not been achieved, although computer science education in the high school is becoming more common. Rather than separate courses in computing, \ken\ envisioned a ``mathematics laboratory" accessible ``to every mathematics student regardless of level." On the practical side,  she anticipated that the reaction of those in the audience might be ``there are no funds, and that takes care of that." She noted, ``This is a situation in which the experiences of life makes me a minor expert."  To convince them, she had contacted potential donors for Catholic schools in the Dubuque area and read aloud letters from some who had responded positively. Regarding teacher training, 
\begin{quote}
The number of such courses that are available at many colleges is yearly increasing, and a single course is sufficient to bootstrap teachers into a program of self-education. ... Education is, after all, a self-activity, and a patient but unrelenting electronic assistant can be a boon.
\end{quote}

\section{SISTER KENNETH, THE PERSON}

\ken\ was witty, fun-loving, and risk-taking.
She had appropriate confidence in her abilities and refused to be intimidated. 
 ``She could not abide fools," \cite{stwo} yet she was generous with her time.
Despite her strong sense of independence, she never learned to drive a car \cite{mone}.
She was curious, creative, and not at all self-centered \cite{bone}.
She was also self-deprecating,
telling stories about embarrassing moments like the time she locked herself out of her hotel room in the middle of the night, mistakingly thinking she was in her own bedroom and had to go down the hall to the bathroom \cite{mone,cthree}.
She also told of a visit with her mother, who had dementia late in life. Her mother asked if \ken\ saw the (non-existent) man outside the window. \ken\ tried to affirm her by saying she did, only to have her mother  respond, ``You know very well I'm not seeing a man out there!" \cite{bone}

\ken\ was very generous; she was thrifty with her limited funds, and if a student or other person needed money, she provided it if she could, very quietly \cite{bone,cthree}.
\ken\ put aside some money for a good party to be held after her death, but one of the sisters turned it in for the general fund instead, disappointing many of \ken's close friends \cite{cone}.

\begin{figure}
\begin{centering}
\includegraphics[width=.99\columnwidth] {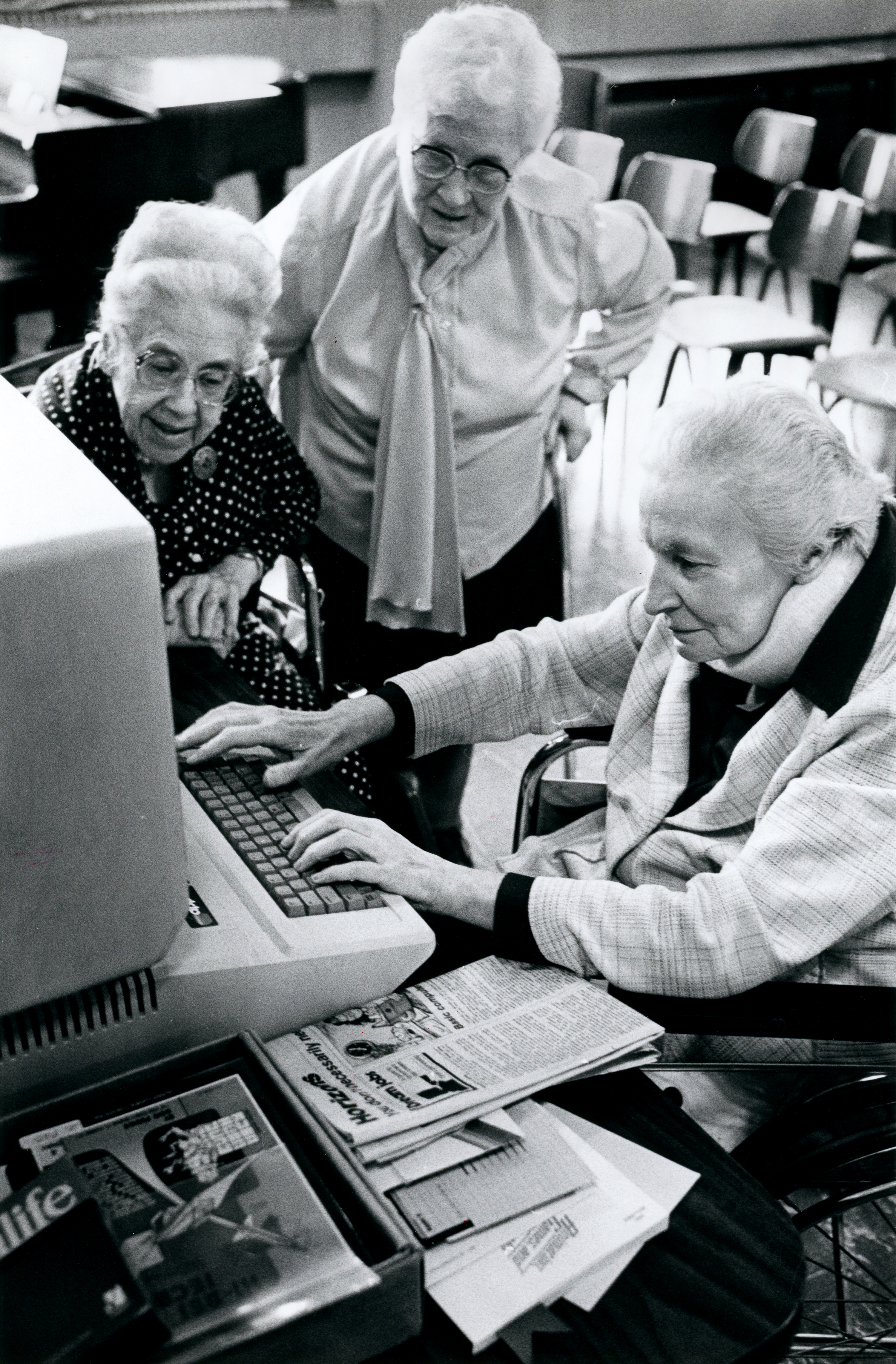}\\
\end{centering}
\caption{\ken\ (on right) in April, 1984, demonstrating computing to BVM sisters Gladys Ramalay and Marian Delaney. Photo Credit: Reprinted with permission of the \emph{Dubuque Telegraph Herald}.}
\label{fig_1984}
\end{figure}

Languages were a ``hobby" to \ken; she could read French, German, Italian, and Russian \cite{c65x}.

The arts were very important to \ken.
She read novels and histories \cite{cone}.
She loved theater and saw many plays twice \cite{cone}.
As an adult, she once acted in a play and ``she {\em was} Gertrude Stein!" \cite{sone} 
She also loved visual arts and was ``very intelligent and perceptive." \cite{cone}
Her interest in movies got her interested in the Alpha Centauri computer game \cite{bone}.
She also loved classical music and enjoyed dancing \cite{cone} and playing violin \cite{eug}.
``Her human spirit was never stifled by her great scientific interests in the machine." \cite{eug}

\ken\ had a keen sense of mission. A sign in her office read, ``My life is a continuing changing awareness of God's will for me." \cite{witn75} She was also a private person, remarking that ``her prayer group was the Father, Son, and Holy Spirit." \cite{eug}
 She kept a handwritten copy of Ecclesiastes 3:1-2:
``There is an appointed time for everything under the heavens. A time to be born, a time to die, a time to plant, and a time to uproot the plant."

\ken\ had a keen sense of humor. She was often recruited by phone to job openings around the country, and she would politely listen to the pitch. When the topic of salary came up, she would surprise the recruiter by saying, ``You know, I couldn't accept a salary since I've taken the vow of poverty." \cite{tele65} 

As part of the scholarship dinner marking the 20th anniversary of Clarke's Computer Science Department, there was a  ``roast in memory of Sister Mary Kenneth. Proceeds from the dinner will be used to establish a scholarship in her name." \cite{c85,scholarship}
``It is the feeling of her colleagues that no one would appreciate a roast more than Sister Kenneth," Sister Margaret O'Brien, SP said \cite{c85}.

\section{FINAL DAYS}

\ken\ spent her final days at Mount Carmel, the Motherhouse in Dubuque, where medical and hospice care was provided for sisters who needed it. She had a computer in her bedroom even as she was dying, and she continued to work \cite{cthree}, even giving lessons to other sisters there ({\bf Figure \ref{fig_1984}}) \cite{c84a} and planning nutritious meals for sisters  \cite{bvm03}.
\cthree\ remembers that \ken\ did not want to die: she felt she still had work to do. Nevertheless, \bone\ reports that her last words were, ``Yes, yes." She died January 10, 1985.

Her gravestone at Mount Carmel, shown in {\bf Figure \ref{fig_grave}}, is small, flat to the ground, and shared with another sister. It has her name and the three most important dates of her life (birth, entrance into the BVM congregation, and death): ``Sister Mary Kenneth Keller, B.V.M.~1913 1932 1985".

\begin{figure}[h]
\begin{centering}
\includegraphics[width=.99\columnwidth]{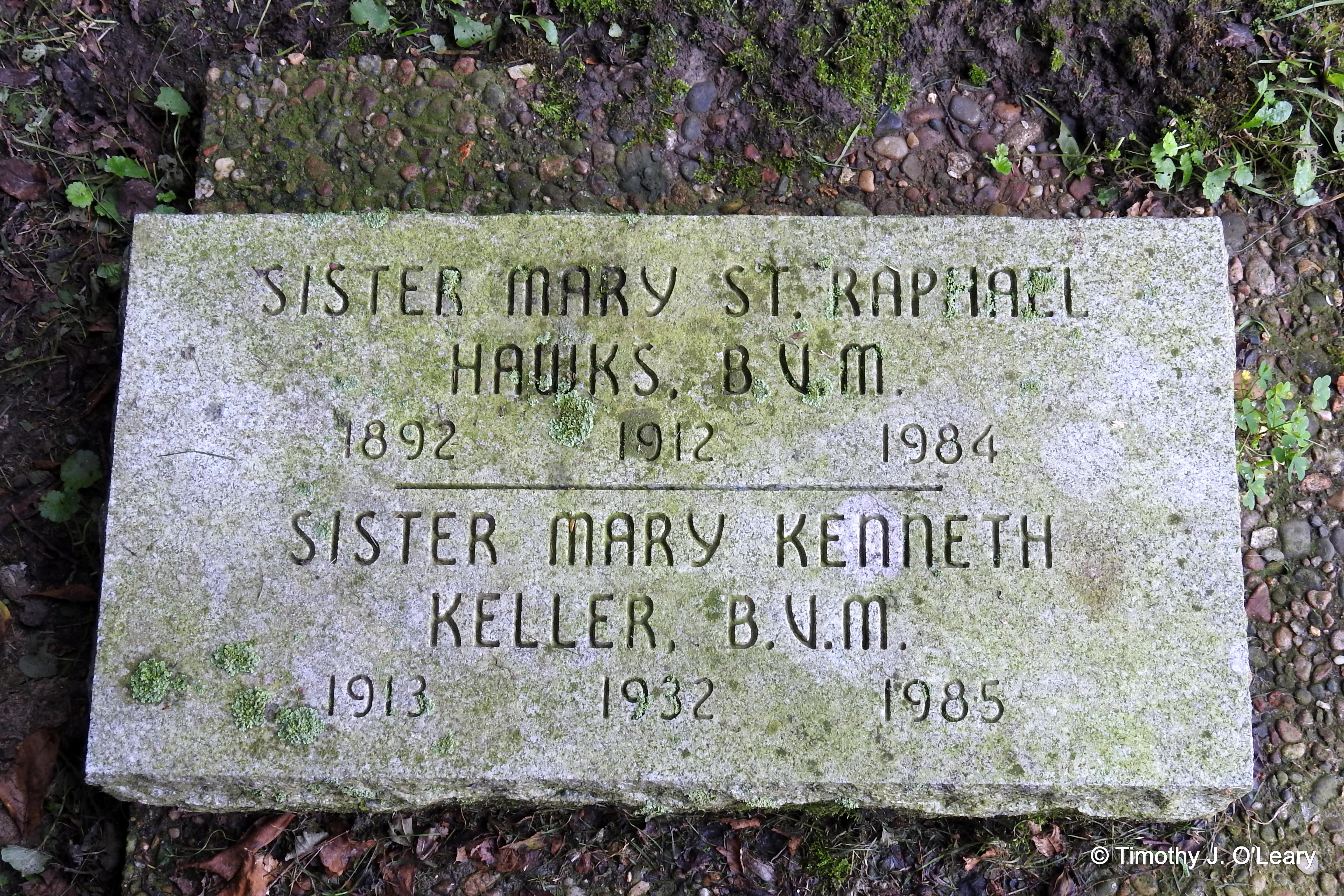}\\
\end{centering}
\caption{Shared gravestone at Mount Carmel, Dubuque, Iowa. Photo: T.J. O'Leary}
\label{fig_grave}
\end{figure}

\section{CONCLUSION}
\ken\ was a totally remarkable woman.
She made the most of every opportunity given to her, even her unlikely path to her late-in-life PhD.
As a scholar, her strength was not in research, but she has the distinction of being an early advocate of learning-by-example in artificial intelligence.
Her main scholarly contribution was in shaping computer science education in high schools and small colleges.
She was an evangelist for the importance of computer literacy for everyone, not just for those in science and technology.
She was far ahead of her time in working to ensure a place for women in technology and in eliminating barriers preventing their participation, such as poor access to education, daycare, and nursing rooms.
She was a visionary in seeing how computers would revolutionize our lives.
She was a strong and spirited woman whose decisions were based on her perception of God's will for her.
Her impact on Clarke University and beyond is due to her hard work, her dedication to education, and her deep sense of mission.

\section*{ACKNOWLEDGMENT}
This work depended upon the help of many people.
We are grateful to 
Carol Blitgen, BVM,
Mary Lou Caffery, BVM,
Catherine Dunn, BVM,
Bertha Fox, BVM,
Diana Malone, BVM,
Sheila O'Brien, BVM,
and
Carmelle Zserdin, BVM
for sharing their memories of \ken, and to Sara McAlpin, BVM, for her memories and her careful reading of the manuscript.

DPO became involved in this work due to publications by 
Ralph London, who was especially generous in his email correspondence, and she was astounded to later discover that
she graduated from the same high school as \ken.


Thomas Byrne,
Robert Hargraves,
Arthur Luehrmann, 
Tracy Flynn Moloney,
and especially Steven Garland, Anthony Knapp, and Thomas Kurtz
helped to shed light on \ken's connection to Dartmouth.

Mary Tonoli Teister (Immaculata Alumnae President),
Mary C. Beckman, BVM (Immaculata Alumnae Office), Marilyn Ferdinand (DePaul University), David Hemmendinger, and Michael Steuerwalt (NSF) also provided useful assistance.

We are especially grateful to Peter J. Murdza Jr., who searched Dartmouth archives for us,
and Karen L. Barrett-Wilt who did the same at the University of Wisconsin.

\bigskip

\section*{About the Authors}

{\bf Jennifer Head} (jhead@bvmsisters.org) is the Archivist for the Sisters of Charity of the Blessed Virgin Mary in Dubuque, IA, a position she has held for 10 years. She holds a BA in History from DePaul University, an MA in American History from the Catholic University of America and a master's degree in Library and Information Science from Dominican University. Jennifer is a Certified Archivist and frequently presents at archival and historical conferences.

{\bf Dianne P. O'Leary} (oleary@umd.edu) is a Distinguished University Professor, emerita, at the University of Maryland, College Park, Maryland. She held appointments in the university's Computer Science Department, Institute for Advanced Computer Studies (UMIACS), and Applied Mathematics \& Statistics and Scientific Computing Program.  She earned a B.S. from Purdue University and a Ph.D from Stanford University.  She has authored over 100 research publications on numerical analysis and computational science, 30 publications on education and mentoring and 17 on the history of scientific computing.  She is a member of AWM and a Fellow of SIAM and ACM.  Further information about her work can be found at \url{http://www.cs.umd.edu/users/oleary}. 
\clearpage

\section{APPENDIX:  SISTER KENNETH'S WORK ON HER HIGH SCHOOL NEWSPAPER}

\bigskip

\begin{centering}
\includegraphics[width=.6\columnwidth]{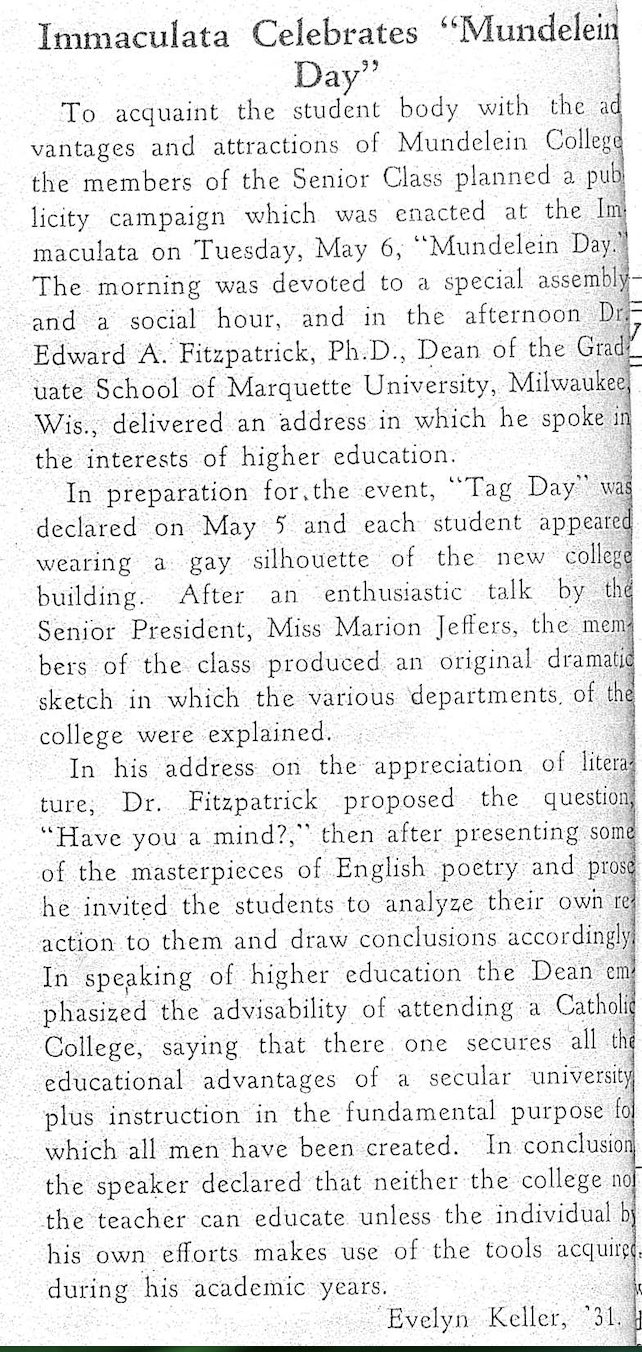}\\
{\emph{The Immaculata News} article authored by \ken\ on ``Immaculata Celebrates `Mundelein Day,'" Vol. 3, Num. 7, p.4.}\\
\end{centering}

\begin{centering}
\includegraphics[width=.6\columnwidth]{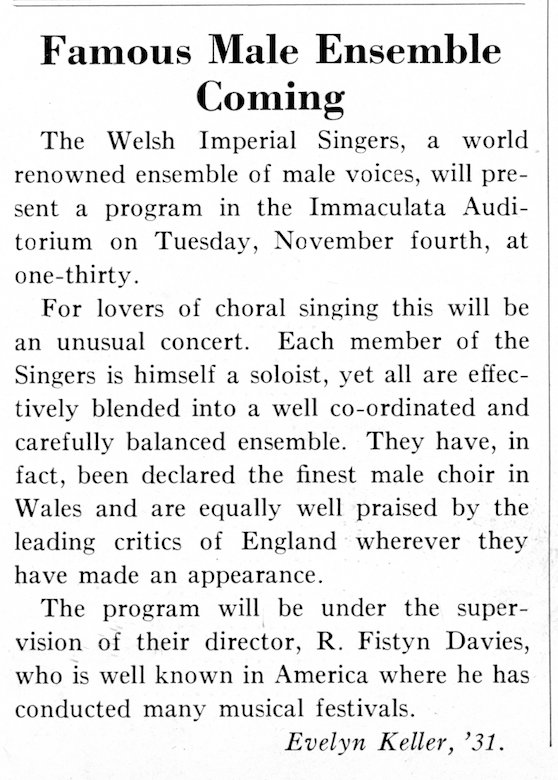}\\
{\emph{The Immaculata News} article authored by \ken\ on ``Famous Male Ensemble Coming," May 1930, Vol. 4, Num. 1, p.3}\\
\end{centering}

\begin{centering}
\includegraphics[width=.6\columnwidth]{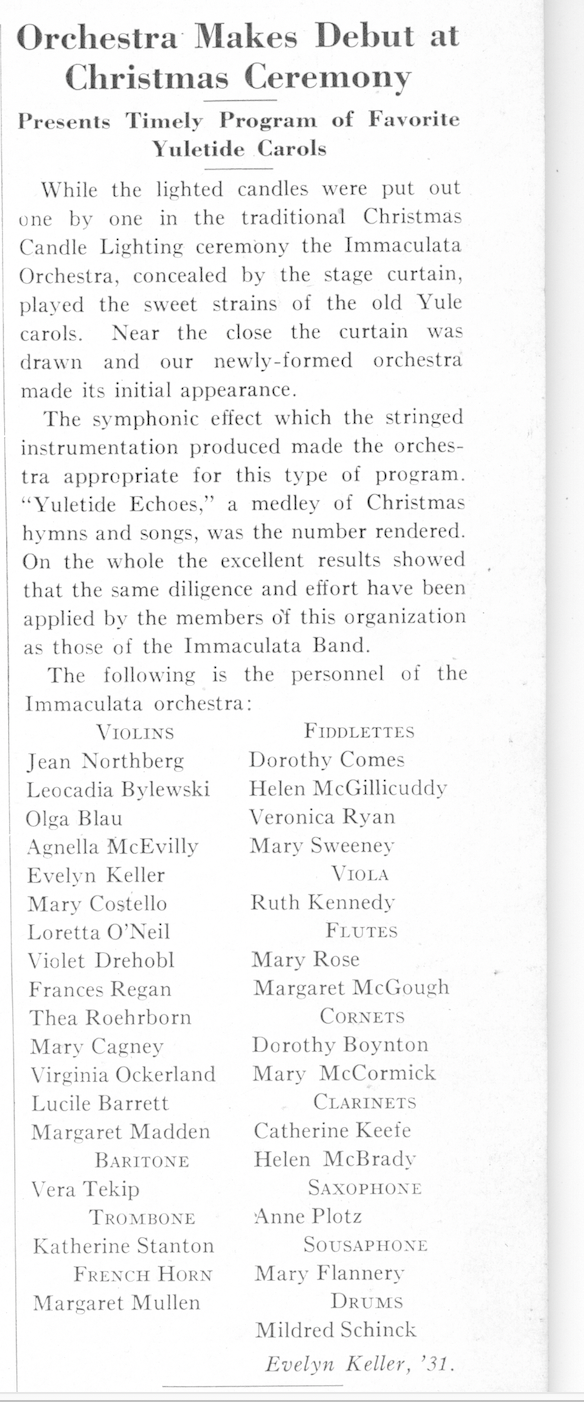}\\
{\emph{The Immaculata News} article authored by \ken\ on, ``Orchestra Makes Debut at Christmas Ceremony," December 1930, Vol. 4, Num. 2, p.7.}\\
\end{centering}

\begin{centering}
\includegraphics[width=.99\columnwidth]{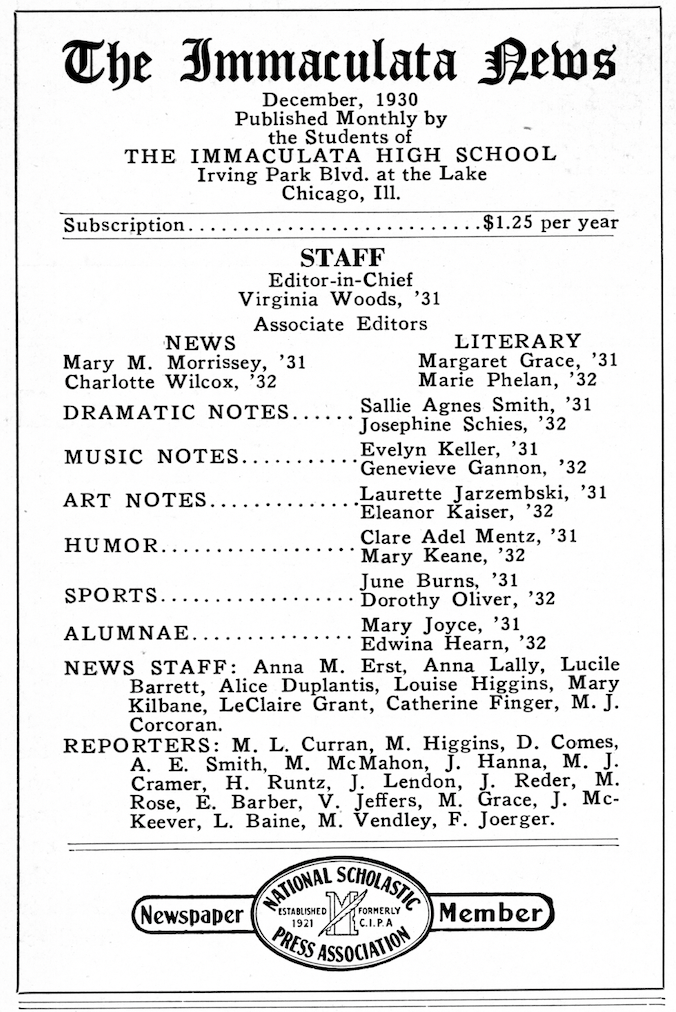}\\
{\emph{The Immaculata News} masthead, December 1930, Vol. 4, Num. 2, p.2.}\\
\end{centering}

\bibliography{ref}
\bibliographystyle{plain}

\end{document}